\documentclass[12pt]{iopart}

\usepackage{graphicx}
\usepackage{cite}
\usepackage[caption=false]{subfig}
\usepackage{amssymb}

\usepackage{xcolor}

\usepackage[UKenglish]{babel}

\usepackage{lineno}

\begin{document}
\setlength{\mathindent}{60pt}

\title[Thermodynamics of Particle Mixtures -- SPT and MC 
]{Thermodynamics of Hard Sphere and Spherocylinder Mixtures – Scaled Particle Theory and Monte Carlo Simulations}

\author{Volodymyr Shmotolokha}
\address{Laboratory of Physical Chemistry, Department of Chemical Engineering and Chemistry \& Institute for Complex Molecular Systems, Eindhoven University of Technology, P.O. Box 513, 5600 MB Eindhoven, The Netherlands}

\author{Jonas Maier-Borst}
\address{Institute of Physics, University of Freiburg, Hermann-Herder-Str. 3, 79104 Freiburg, Germany}

\author{Mark Vis}
\address{Laboratory of Physical Chemistry, Department of Chemical Engineering and Chemistry \& Institute for Complex Molecular Systems, Eindhoven University of Technology, P.O. Box 513, 5600 MB Eindhoven, The Netherlands}

\author{Anja Kuhnhold$^{{\ast},\#}$}
\address{Institute of Physics, University of Freiburg, Hermann-Herder-Str. 3, 79104 Freiburg, Germany}
\ead{anja.kuhnhold@physik.uni-freiburg.de}

\author{Remco Tuinier$^{{\ast},\#}$}
\address{Laboratory of Physical Chemistry, Department of Chemical Engineering and Chemistry \& Institute for Complex Molecular Systems, Eindhoven University of Technology, P.O. Box 513, 5600 MB Eindhoven, The Netherlands}
\ead{r.tuinier@tue.nl}

\bigskip

\noindent $^{\ast}$ corresponding authors  \\
$^\#$ contributed equally

\vspace{10pt}
\begin{indented}

\vspace{10pt}
\item[]
\end{indented}

\clearpage
\begin{abstract}
We review the literature on scaled particle theory (SPT) and its extensions and discuss results applied to describe the thermodynamics of hard particle mixtures. After explaining the basic concepts of scaled particle theory to compute the free energy of immersing a particle into a mixture, examples are discussed for the simple case of a hard sphere dispersion and the free volume fraction of ghost spheres in a hard sphere dispersion. Next, the concept is applied to mixtures, and general expressions are shown that relate the free volume fraction in mixtures to the key thermodynamic properties, such as the chemical potential(s) and (osmotic) pressure. Subsequently, it is revealed how these concepts can be extended towards multi-component systems. It is shown that free volume fractions provide chemical potentials and total pressure of multi-component mixtures, and thereby yield the full equation of state. We present novel results for ternary particle dispersions composed of hard spherocylinders and two types of hard spheres differing in size. Throughout, we show the accuracy of SPT by comparing the results with those of Monte Carlo computer simulations.
\end{abstract}

\vspace{2pc}

\maketitle
%

\section{Introduction}
Scaled Particle Theory (SPT) was developed to derive expressions for the thermodynamic properties of particle fluids by relating them to the reversible work needed to insert an additional particle into the system. The basis of SPT is to calculate an expression for the work of insertion $W$ \cite{reiss1959}, pioneered by Reiss, Frisch, and Lebowitz in 1959. SPT turned out to be a powerful tool for determining the work required to insert an additional particle into a fluid, in order to predict the thermodynamic properties of fluids.

 Reiss and colleagues further expanded on these concepts, introducing scaled particle methods and elegantly linking these to the statistical thermodynamics of real fluids~\cite{reiss1965,reiss1960,helfand1961}. Subsequently, Scaled Particle Theory (SPT) became a highly successful tool in the theory of liquid state physics \cite{stillinger1973,ashbaugh2006}. SPT is not exact, but it is relatively simple, often accurate, and provides algebraic expressions for thermodynamic quantities. Stillinger and Lebowitz and their collaborators~\cite{lebowitz1965,stillinger1973} made significant steps in extensions of SPT, examining structures in solutions and fluid mixtures, respectively. {Tully-Smith and Reiss}~\cite{tullysmith1970} and {Mandell and Reiss}~\cite{mandell1975} focused on the development and application of SPT for rigid sphere fluids. {Bergmann} and collaborators~\cite{bergmann1976,tenne1978} extended SPT to non-additive hard spheres and provided solutions to include positive non-additivity in SPT.

In the 1970s, Cotter and colleagues \cite{Cotter1970,Cotter1972,Cotter1974,Cotter1977} developed the SPT approach to describe the thermodynamic properties of a fluid composed of anisotropic particles. Cotter and Stillinger \cite{Cotter1972} delved into two-dimensional systems and extended SPT to rigid disks. Cotter \etal \cite{Cotter1970,Cotter1974,Cotter1977} used SPT to study the thermodynamic properties of fluids of rod-like particles. Cotter \cite{Cotter1974,Cotter1977} derived SPT expressions for a collection of hard spherocylinders. Whereas for infinitely long thin hard rods, Onsager's \cite{Onsager1949} second virial theory is exact, it deviates for shorter rods or attractive rods. Cotter's approach is an approximate way to predict the equation of state to take into account higher virial coefficients. In Refs.~\cite{PetersPRE2020} and \cite{lekkerkerker_tuinier_vis_2024} SPT predictions for hard spherocylinders were compared to computer simulations, and close agreement was found. 

When SPT is applied to colloidal dispersions, the particles are usually approximated as hard particles, meaning they only interact through their excluded volumes: they cannot overlap, and there are no additional interactions. The SPT concepts have been applied to binary colloidal mixtures but were first applied to colloid-polymer mixtures \cite{Lekkerkerker1990} in the context of the so-called free volume theory (FVT). In FVT, the added particles are described via an osmotic reservoir that is in contact with the system of interest. The advantage of such an osmotic equilibrium approach is that one can fix the fugacity (or, likewise, the chemical potential) in the system by imposing a certain fugacity of the added particles in the reservoir. In classical FVT \cite{Lekkerkerker1992}, however, these added particles are treated as behaving ideally. For colloid-polymer mixtures, this approach has been extended to describe rod-polymer \cite{Lekkerkerker1994,Tuinier2007}, platelet-polymer \cite{AGG2017}, and cube-polymer mixtures \cite{AGGCubes2018}. It is also possible within this approach to reasonably accurately describe the interactions between the polymers \cite{Aarts2002,Peters2022JPCM} or to account for charged colloids \cite{Fortini2005}. 

Binary asymmetric hard sphere {(HS)} mixtures were studied using FVT as well \cite{Lekkerkerker1993}, guided by the work on colloid--polymer mixtures \cite{Lekkerkerker1992}. The work of Lekkerkerker and Stroobants \cite{Lekkerkerker1993} suggests that the addition of small hard colloidal spheres to a mixture of large hard spheres will lead to a fluid--solid phase separation and that there is a metastable gas--liquid phase separation. Later, this finding of a metastable gas--liquid phase coexistence was confirmed by computer simulations of Dijkstra, van Roij and Evans \cite{Dijkstra1999a}. Opdam \etal \cite{Opdam2021} proposed a more rigorous approach to account for the interactions between the particles in the reservoir. For asymmetric {HS} mixtures, the FVT predictions of Opdam \etal \cite{Opdam2021} are much closer to the computer simulation results of Ref.~\citenum{Dijkstra1999a}. {Koda and Ikeda \cite{Koda2002} made a computer simulation study of a system of parallel hard spherocylinders (HSC) and found semi-quantitative agreement between SPT and their simulation results. Hvodz \etal \cite{Hvozd2018} studied isotropic–nematic phase equilibria for the bulk HS/HSC mixture and compared these with computer simulations.}

{An early SPT extension was the application to hard chain fluid mixtures, where SPT was used to improve equations of state and chemical potentials \cite{omelyan2001self,okeefe2017scaled}.} Gibbons \cite{Gibbons1969} derived SPT expressions for the thermodynamic properties of arbitrary shapes, which can, for instance, also be used to describe dispersions of hard superballs to mimic colloidal cubes \cite{lekkerkerker_tuinier_vis_2024}. {Boubl{\'{i}k \cite{Boublik1975} derived a general expression for the equation of state of a mixture of different types of hard convex bodies using SPT. This approach was later extended to porous media systems to take into account confinement effects on thermodynamic properties \cite{HOLOVKO201430}.}}

Scaled Particle Theory (SPT) has found widespread utility in modelling a variety of complex fluids beyond hard particles and has been applied to a large number of systems of which we give an overview below. 

{Ashbaugh and Pratt}~\cite{ashbaugh2006} discussed the length scales of hydrophobicity in the context of SPT. The mathematical and numerical intricacies of SPT for hard sphere pairs were explored by {Stillinger, Debenedetti, and Chatterjee}~\cite{stillinger2006, chatterjee2006}. {Heying and Corti}~\cite{heying2004,heying2014} revisited and enhanced the SPT, introducing new conditions and improved predictions for hard-sphere fluids. {Siderius and Corti}~\cite{siderius2006, siderius2005a} significantly adapted the theory, focusing on hard-sphere equations of state, multiple interpolation functions, and inhomogeneous hard particle fluids. SPT was employed to predict surface tension of binary \cite{latifi2010extended} and ternary mixtures \cite{mohsennia2011measurement}.

SPT has been adapted to describe anisotropic and confined systems. It was utilised to investigate the isotropic–nematic phase transition in mixtures of hard spheres and spherocylinders \cite{holovko2017isotropic}. Later, it was extended to describe hard sphere fluids in random porous matrices, capturing confinement-induced shifts in thermodynamic behaviour for both single- and multi-component systems \cite{dong2018scaled}.

Applications of SPT to ionic liquids gained traction with studies focused on thermodynamic properties, solubility, and volumetric behaviour \cite{zafarani2015study,shekaari2016thermodynamics,akbari2018solubility,zafarani2019effect}. These were followed by detailed solvation studies where SPT was used to decompose the standard partial molar volume into cavity formation and solute–solvent interaction contributions. Notably, this was applied to ionic liquids in ternary salt mixtures \cite{shekaari2019solvation}, and to pharmaceutical solutes such as paracetamol in amino acid-based ionic liquids \cite{shekaari2021paracetamol}.

Deep eutectic solvents (DESs) emerged as another promising domain for SPT applications. Karimi \etal~\cite{karimi2020sweetness} studied glucose and fructose in aqueous DESs, using SPT to quantify cavity and interaction volumes under varying concentrations and temperatures. Similar work explored solute behaviour in different DES formulations, highlighting how solvent composition affects hydration and solvation efficiency \cite{behboudi2020effect,mohammadi2021study}.

SPT has also been instrumental in the study of solvation free energies. For example, standard Gibbs energies and excess Henry’s constants for nonpolar gases in complex aqueous mixtures were modelled using SPT \cite{mainar2019solubility}. More recently, SPT was integrated into continuum solvation models to estimate cavitation energies, yielding good agreement with experimental solvation data \cite{minenkov2023solv}. It was also used to compute the solvation free energy by combining solvent-accessible volume and area models \cite{akkus2023revisiting}.

In parallel, the theory has been increasingly applied to soft matter and biological systems. It was used to model phase transitions in star polymers \cite{hasegawa2019lyotropic} and the isotropic–nematic phase separation of rod-like amylose derivatives \cite{kim2019lyotropic}. Crowding effects in protein systems were studied with hybrid SPT-Fundamental Measure Theory \cite{qin2010generalized}, later extended to dimerisation behaviour in proteins under macromolecular crowding \cite{pradhan2023gb1}, and to the stability of protein-genome complexes in crowded media \cite{delgado2023crowding}. Volumetric changes during ligand binding to the SARS-CoV-2 protease were also quantified using SPT \cite{alvarado2022interaction}.

The geometrical framework of SPT has undergone significant evolution. A major advancement was an extension which incorporates higher-order curvature terms to improve the prediction of surface tension and bending rigidity near curved interfaces \cite{qiao2020augmented,qiao2022augmented}. Adapted SPT was proposed to analyse interfacial properties of hard-disk fluids, with improved accuracy over classical SPT \cite{martin2020surface}. Additional refinements yielded closed-form expressions for structural correlations in two-dimensional systems \cite{heying2021predicting}, and were validated with simulation data.

Vuorte \etal \cite{vuorte2023equilibrium} used SPT to investigate crowding among surface-adsorbed surfactant aggregates, treating them as hard disks and linking bulk aggregation with surface packing. The theory has also been incorporated into the modelling of charged colloidal systems; for example, SPT was combined with theory for the effective diameter of charged rods \cite{SLO1986} to predict osmotic pressure and free volume in charged spherocylinder dispersions \cite{tuinier2023equation}.

Holovko, Dong, and Shmotolokha~\cite{holovko2009, holovko2010} derived analytical SPT for hard sphere fluids in random porous media. SPT turned out to be particularly useful for describing mixtures in bulk and in the presence of porous media, where the size and shape of the particles can significantly influence the thermodynamic properties. Patsahan, Holovko and Dong \cite{Patsahan2011} proposed a novel formulation in the basic SPT formulation for hard-sphere fluids in a hard-sphere matrix, which is in close agreement with computer simulation results. These results were further improved to also correctly describe results at large volume fractions \cite{Holovko2017}. Chen \etal\cite{Chen2016} developed SPT for multi-component HS fluids confined in a multi-component HS or multi-component overlapping HS matrix. The accuracy of various extensions derived from the basic SPT formulation was evaluated against simulation results.

The impact of particle anisotropy and the porous structure on phase transitions was also studied using SPT \cite{holovko2015, shmotolokha2022, holovko2018, holovko2020}. This allows for a better understanding of the stability and phase equilibrium of such complex systems and opens up new possibilities for their modelling. Hvozd \etal\cite{Hvozd2018} developed SPT to quantify the thermodynamics and orientational ordering in a HS/HSC mixture in disordered porous media. Holovko and co-workers further used the SPT developed for HS/HSC binary mixtures to study confined mixtures. In Refs.~[\citenum{Hvozd2022JML}] and [\citenum{Hvozd2025}], the results of \cite{Hvozd2018} were used to study the phase behaviour of ionic liquids in disordered porous media. 

{In summary, across a wide range of chemical systems -- from hard chain fluids and hydrocarbon mixtures to ionic liquids and biological macromolecules -- SPT consistently provides a robust theoretical framework for predicting thermodynamic and interfacial properties. Its adaptability and predictive power make it a valuable tool in both fundamental research and practical applications. We further focus now on rod-sphere mixtures.}

{Early approaches of FVT on rod-sphere mixtures \cite{vliegenthart1999, oversteegen2005} implicitly assumed the rods in the reservoir behave ideally.} The approach of Opdam \etal was also extended to hard spheres mixed with hard spherocylinders \cite{Opdam2021b} and appeared to describe experiments much better. Also, it turned out this method of Opdam \etal is in close agreement with computer simulations \cite{Opdam2022}. More recently, Opdam \etal \cite{Opdam2023} extended this approach to include higher-order phase states and predicted phase diagrams of rod-sphere, rod-platelet, and sphere-platelet mixtures, demonstrating a very rich phase behaviour.

In this review, we focus on the work of inserting a particle into a colloidal dispersion as predicted using SPT. This directly provides the free volume fraction in a system. It is shown how and that this provides the key thermodynamic properties chemical potential (directly linked to the fugacity) and the (osmotic) pressure of a fluid. We start by focusing on dispersions of a single type of colloidal particle in a (background) solvent. Next, we move on to multi-component colloidal mixtures and derive general expressions that relate the chemical potential and pressure to the free volume fraction. {Throughout, we compare to computer simulation results.}

{SPT has been applied to single and two-component mixtures. One may wonder whether SPT is accurate for a three-component mixture. In this work, we therefore evaluate whether SPT is accurate for three-component mixtures by comparison of the predicted pressure of the ternary mixture with Monte Carlo computer simulations.} {We} show new results for a ternary particle mixture of an asymmetric hard sphere mixture plus added hard spherocylinders. {Such ternary mixtures may also serve as minimal models or reference systems for studies of phase behaviour and can be extended to more complex systems.} 

\clearpage
\section{Classical examples of SPT results}\label{sec:classic}
Here, we illustrate the power of the SPT approach for two examples {from Ref.~\cite{lekkerkerker_tuinier_vis_2024}} related to hard spheres. First, we show how SPT yields the pressure of a hard sphere dispersion. Secondly, it is demonstrated that the free volume fraction available for ghost spheres in a hard sphere dispersion can accurately describe computer simulations of the free volume fraction of ideal polymer chains in a dispersion of hard spheres.

\subsection{Pressure from SPT via the work of insertion into a hard sphere fluid}

Consider $N_1$ hard spheres with diameter $d_1$ and volume $v_1 = \pi d_1^{3}/6$ in a total volume $V$ to which $N_2$ particles 2 are added in a background solvent. The chemical potential $\mu_2$ of a particle 2 in the hard sphere fluid with volume fraction $\phi_1=v_1 N_1/V$ is given by:
\begin{equation} \label{Widomeq}
    \mu_2 = \mu_2^{*} + k_\mathrm{B}T \ln \frac{N_2}{V} + W,
\end{equation} 
where $\mu_2^{*}$ is the reference chemical potential and $W$ is the reversible work required for inserting a particle 2 in a collection of particles 1. 

Widom's particle insertion theorem \cite{Widom1963} states that we can also write:
\begin{equation} \label{Widomeqmughost2}
    \mu_2 = \mu_2^{*} + k_\mathrm{B}T \ln \frac{N_2}{\langle V_\mathrm{free} \rangle _1 } ,
\end{equation} 
where $\langle V_\mathrm{free} \rangle _1 $ is the ensemble-averaged free volume available for particles 2 in a collection of particles 1. We define the free volume fraction for particles 2 in a dispersion of particles 1 as:
\begin{equation} \label{alfafreevol}
 \alpha= \frac{\langle V_\mathrm{free} \rangle _1 }{V} .
\end{equation} 

\begin{figure}[ht]
  \centering
  \includegraphics[width=0.4\linewidth]{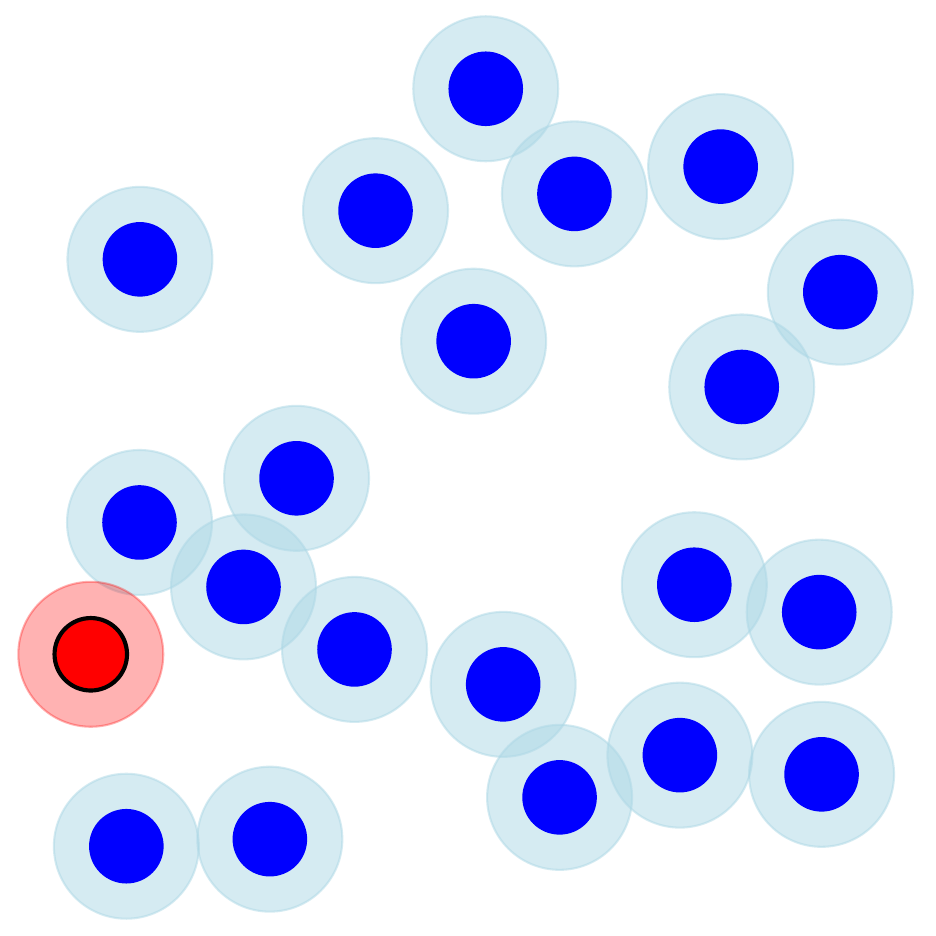}
  \caption{
  Illustration of the free volume fraction $\alpha$ in a single-component particle dispersion; it equals the available volume (total volume minus volume occupied by particles and their excluded volume halos) divided by the total volume. The red particle can only be inserted in the white regions; otherwise, it overlaps with the blue particles.
  }
  \label{single_particle_dispersion}
\end{figure}

Figure~\ref{single_particle_dispersion} illustrates the concept of the free volume fraction $\alpha_1$ available in a system containing only one type of particle. Blue circles represent existing particles with their excluded volume regions (light blue areas). The red circle is a test particle inserted into the system. The centre of the red particle can only be placed in the volume outside the blue areas, including their excluded volume halos.

From Eqs.~(\ref{Widomeq}), (\ref{Widomeqmughost2}) and (\ref{alfafreevol}) it follows that the free volume fraction $\alpha$ available for a particle 2 in a collection of particles 1 is related to the free energy of immersing a particle 2, $W$ as:
\begin{equation}
\alpha = \mathrm{e}^{-{\beta W}}.
\label{eq:alfaW}
\end{equation}
Here $k_\mathrm{B}T= \beta^{-1}$ is the thermal energy. 

\begin{figure}[ht]
  \centering
  \includegraphics[width=0.6\linewidth]{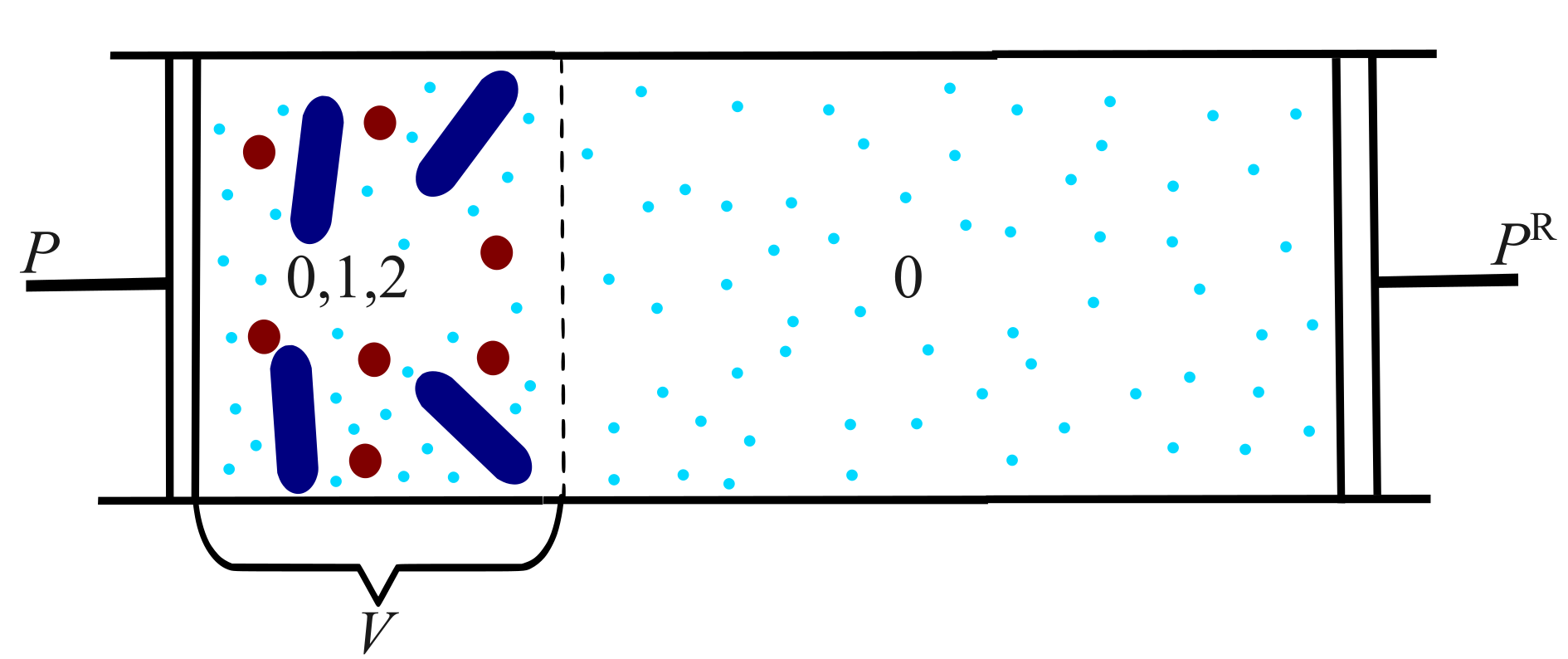}
  \caption{
  {Osmotic equilibrium between a system of interest (left) containing (as an example) particles 0 (solvent), 1 (spheres) and 2 (rods) in contact with a reservoir (right) that only contains solvent. The system has a volume $V$ and pressure $P$. The reservoir has a pressure $P^\mathrm{R}$ and a volume $\gg V$. The separating membrane (dashed) is impermeable to components 1 and 2, but allows permeation of solvent (component 0). The osmotic pressure $\Pi= P- P^\mathrm{R}$.}
  }
  \label{osmpress}
\end{figure}

The work $W$ can be calculated using SPT by scaling the size of the sphere to be inserted: the size of the scaled particle is $\lambda \sigma$, with $\lambda$ being the scaling parameter. In the limit $\lambda \rightarrow 0$, the inserted sphere approaches a point particle, and it is highly unlikely that this particle overlaps with others. In such a case, the free volume fraction can be written as
\begin{equation}\label{alfasmalllambda}
    \alpha  =    1- \phi_1 \left(1 + \frac{\lambda \sigma}{d_1}\right)^3.
\end{equation}
Hence it follows from Eq.~(\ref{eq:alfaW}) that:
\begin{equation} \label{Wsmalllambda}
    \beta W = - \ln \left[ 1- \phi_1 \left(1 + \frac{\lambda \sigma}{d_1}\right)^3 \right]  
\end{equation}
for $\lambda \ll 1$. In the opposite limit $\lambda \gg 1$, when the size of the
inserted scaled particle is very large, $W$ (to a good
approximation) is equal to the volume work needed to create a
cavity $\frac{\pi}{6} (\lambda \sigma)^3$ and is given by
\begin{equation} \label{Wbiglambda}
    W = \frac{\pi}{6} (\lambda \sigma)^3 \Pi_1 ,
\end{equation}
where $\Pi_1$ is the (osmotic) pressure of the hard sphere dispersion. Since we have colloidal dispersions in mind, we use the term osmotic pressure in this paper, but the results derived hold similarly for the pressure $P$ of a collection of particles. {The concept of osmotic pressure is illustrated in Fig.~\ref{osmpress}. Without solvent, $P^\mathrm{R}$ = 0.} 

In SPT\index{SPT}, the above two limiting cases are connected by expanding $W$ of Eq.~(\ref{Wsmalllambda}) as a Taylor series in $\lambda$ up to $\lambda^2$ and then add Eq.~(\ref{Wbiglambda}) as $\lambda^3$-term:
\begin{equation} \label{Wexpansion}
    W(\lambda) = 
        W(0) + \left( \frac{\partial W}{\partial \lambda}\right)_{\lambda=0} \lambda 
        + \frac{1}{2} \left( \frac{\partial^2 W}{\partial \lambda^2}\right)_{\lambda =0} \lambda^2 
        + \frac{\pi}{6}(\lambda \sigma)^3 \Pi_1.
\end{equation}
Using $q=\sigma/d_1$ this yields for $\lambda=1$:
\begin{equation} \label{Wexpansionlambda1}
    \beta W(\lambda=1) =  
        - \ln (1 - \phi_1)
        + \frac{3q \phi_1
        }{1 - \phi_1}
        + \frac{3q^2 \phi_1}{1 - \phi_1}
        + \frac{9 q^2 \phi_{1}^{2}}{2(1 - \phi_1)^2} 
        + \frac{\pi}{6} q^3 d_1^{3} \beta \Pi_1 .
\end{equation}

The pressure $\Pi_1$ of the pure hard sphere system can now be obtained \cite{reiss1959} from the reversible work of inserting an identical sphere ($\sigma=d_1$). For this situation ($q=1$), Eq.~(\ref{Wexpansionlambda1}) becomes:
\begin{equation} \label{Wexpansion1part}
    \beta {W} = - \ln (1-\phi_1) + 6 \frac{\phi_1}{1- \phi_1} + \frac{9}{2}\left(\frac{ \phi_1}{1 - \phi_1}\right)^2 + \beta \Pi_1 v_1
    = \beta \mu^{\mathrm{ex}}_1
    .
\end{equation}
Using 
\begin{equation} \label{mu1}
   \mu_1 = \mu^{\mathrm{id}}_1 + \mu^{\mathrm{ex}}_1 = \mu^{\mathrm{0}}_1 +  k_\mathrm{B}T \ln \phi_1 + W
    ,
\end{equation} 
with the ideal part for the chemical potential given by $\mu^{\mathrm{id}}_1  = \mu^{\mathrm{0}}_1 +  k_\mathrm{B}T \ln \phi_1$, and applying the Gibbs--Duhem relation for a 1-component system:
\begin{equation}\label{GD}
v_1\frac{\partial  \Pi_1 }{\partial \phi_1}=\phi_1 \frac{\partial 
\mu_1}{\partial \phi_1} 
,
\end{equation}
yields\footnote{For the derivation of Eq.~\ref{PY} it is convenient to use $y_1=\phi_1/(1-\phi_1)$}:
\begin{equation} \label{PY}
    \beta \Pi_1 v_1= \frac{\phi_1 + \phi_1^2 + \phi_1^3}{(1- \phi_1)^3},
\end{equation}
which is the famous SPT expression for the pressure of a hard sphere
fluid \cite{reiss1959}. This expression equals the classical Percus-Yevick compressibility route result \cite{PY1958}, which is in close agreement with computer simulation results up to volume fractions of at least 40 vol\% \cite{Vrij2005}.

\subsection{Free volume fraction of ghost particles in a hard sphere fluid}
\begin{figure}[ht]
  \centering
  \includegraphics[width=0.45\linewidth]{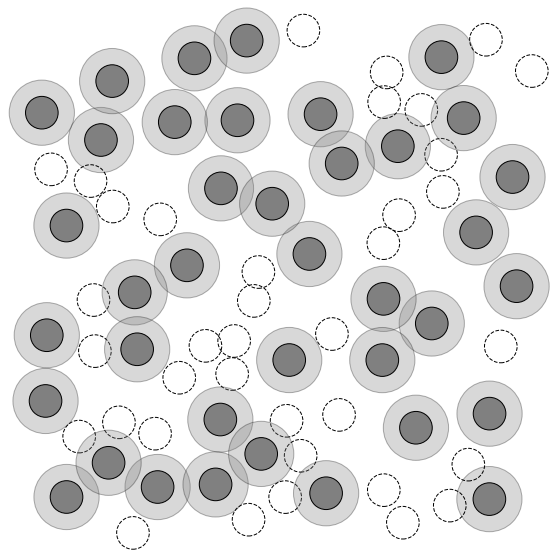}
  \caption{Simple picture of a colloid-polymer mixture: a dispersion of hard spheres (dark grey circles) mixed with ghost particles (indicated by the dashed circles) as a model for ideal polymers. The ghost spheres can freely overlap and behave ideally. The depletion thickness is set by the ghost sphere radius \cite{Eisenriegler1983,lekkerkerker_tuinier_vis_2024}.}
  \label{fig:HSghost}
\end{figure}

Next, we consider a hard sphere fluid of particles 1 and add ghost particles 2, as sketched in Fig.~\ref{fig:HSghost}. These ghost particles, also called penetrable hard spheres \cite{DeHek1981}, do not feel each other and hence behave ideally, but cannot overlap with the hard sphere particles 1. We can use again Eq.~(\ref{Wexpansionlambda1}) and insert Eq.~(\ref{PY}) for the pressure of the hard spheres 1 and write for arbitrary $q$: 
\begin{equation} \label{Wexpansion1partplus2}
    \beta W =  
        - \ln (1 - \phi_1)
        + \frac{(3q+3q^2)\phi_1}{1 - \phi_1}
        + \frac{9 q^2 \phi_{1}^{2}}{2(1 - \phi_1)^2} 
        + q^3 \left(\frac{\phi_1 + \phi_1^2 + \phi_1^3}{(1- \phi_1)^3} \right).
\end{equation}
Insertion of $W$ from Eq.~(\ref{Wexpansion1partplus2}) into Eq.~(\ref{eq:alfaW}) yields the following result for the free volume fraction $\alpha$:
\begin{equation}
\alpha = (1-\phi_1) \mathrm{e}^{-\left[(3+3q+q^2)\frac{q\phi_1}{1-\phi_1}+(\frac{9}{2} + 3 q)\left(\frac{q\phi_1}{1-\phi_1}\right)^2 +3 \left(\frac{q\phi_1}{1-\phi_1}\right)^3  \right]},
\label{eq:alfaW12}
\end{equation}

\begin{figure}[ht]
  \centering
    \includegraphics[width=0.7\linewidth]{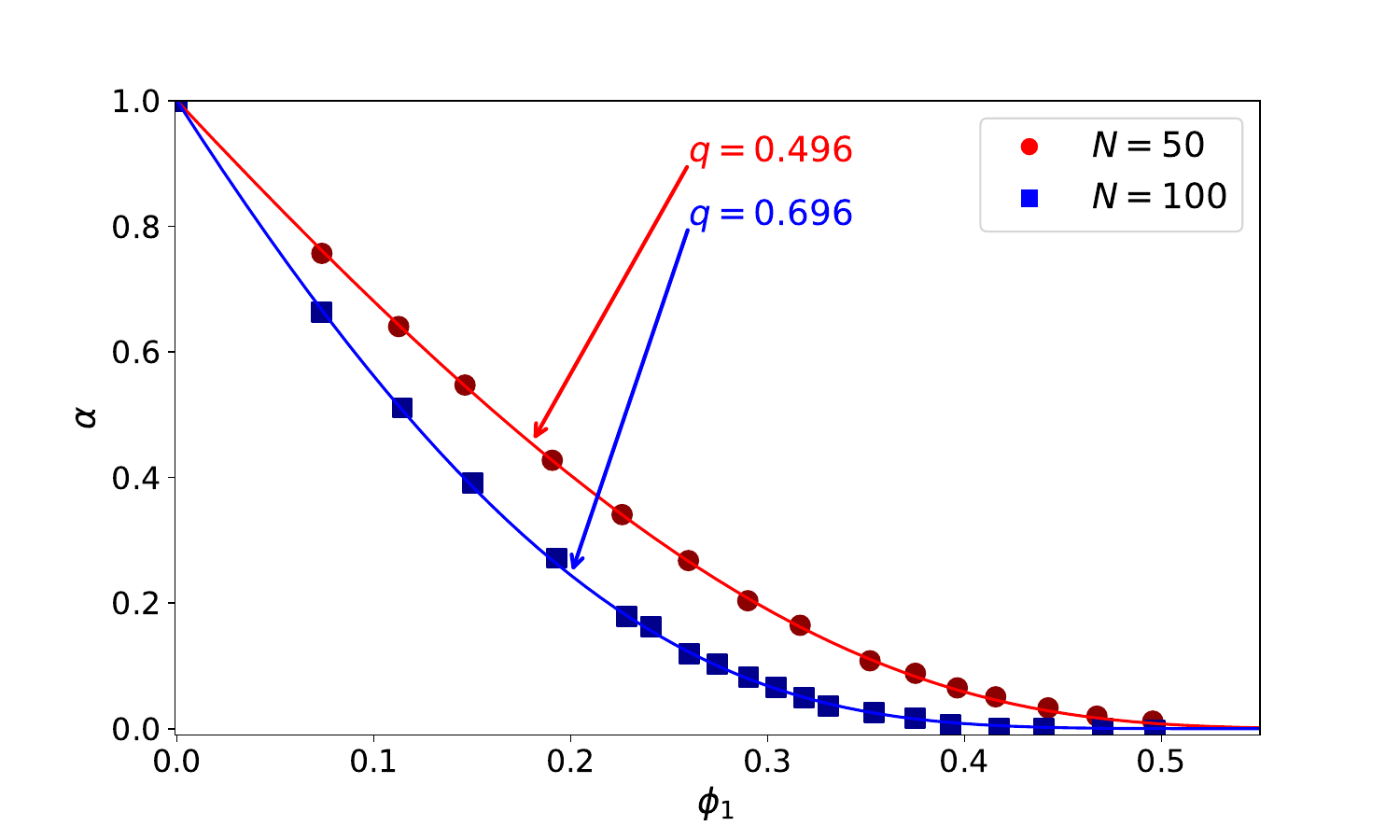}
  \caption{Free volume fraction of ghost spheres (curves; SPT) and ideal polymer chains (symbols; Monte Carlo computer simulations, redrawn from Meijer and Frenkel \cite{Meijer1994}) as a function of the hard sphere volume fraction of particles 1 with diameter $d_1 =$ 10.5 lattice units. The polymer chains were modelled as lattice chains of $N=50$ and 100 segments (see \cite{Meijer1994} for details). Each segment occupies one lattice unit. }
  \label{fig:alfaghostsims}
\end{figure}

In Fig.~\ref{fig:alfaghostsims}, we compare the SPT predictions for the free volume fraction of ghost spheres in a hard sphere dispersion with computer simulations by Meijer and Frenkel \cite{Meijer1994}. They simulated hard spheres mixed with ideal polymer chains. In this case, ghost spheres mimic ideal polymers quite well once the correct size of the ghost spheres is taken, being equal to the depletion diameter. {The effective depletion size can be computed from the excess density profile of the depletant particle next to a surface \cite{lekkerkerker_tuinier_vis_2024}.}

\clearpage  

\section{Exact relations between free volume fraction, chemical potential and pressure}

\subsection{Single particle type dispersion}

We first show that the relative volume available in a system of interest is a key thermodynamic quantity for a dispersion containing only a single type of particle 1 in a background solvent. As in Sec.~\ref{sec:classic} we make again use of Widom's particle insertion theorem \cite{Widom1963}. Hence, we use that the free volume fraction $\alpha_1$ is directly related to the immersion free energy $W$ of inserting particle 1 in a sea of particles 1. The excess chemical potential of the particle $\mu^\mathrm{ex}_\mathrm{1}=k_\mathrm{B}T\widetilde{\mu}^\mathrm{ex}_\mathrm{1}=\mu_{1}-\mu_{1}^{0}-k_\mathrm{B}T \ln\phi_\mathrm{1} $, where $\mu_{1}^{0}$ is again a  (reference) chemical potential, $\widetilde{\mu}=\beta {\mu}$, and $\phi_\mathrm{1}$ is the particle volume fraction of component 1, is related to $W$ through:
\begin{equation}\label{muexW}
\widetilde{\mu}^\mathrm{ex}_\mathrm{1}={\beta W}=-\ln \alpha_{1}(\phi_\mathrm{1}) .
\end{equation}
The Gibbs-Duhem relation (\ref{GD}) can also be written as:
\begin{equation}\label{GD1}
\beta v_1 \frac{\partial  \Pi_\mathrm{1}}{\partial \phi_\mathrm{1}}=
1+ \phi_\mathrm{1} \frac{\partial \widetilde{\mu}^\mathrm{ex}_\mathrm{1}}{\partial \phi_\mathrm{1}}
,
\end{equation}
Integration and using Eq.~(\ref{muexW}) provides \cite{TuinierKuhnhold2023}:
\begin{equation}\label{EOSalfa}
\beta \Pi_\mathrm{1} v_1=\phi_\mathrm{1}-\phi_\mathrm{1}\ln \alpha{_1}(\phi_\mathrm{1}) + \int_{0}^{\phi_\mathrm{1}} \ln \alpha_{1}(\phi_\mathrm{1}') \mathrm{d}\phi_\mathrm{1}'
,
\end{equation}
which gives a direct relation \cite{Vortler2000} between the free volume fraction available for a particle in the system and the (osmotic) pressure $\Pi_1$. Hence, it turns out $\alpha$ yields the equation of state.

\subsection{Binary particle dispersion}

The Gibbs-Duhem relation for a multi-component mixture reads:
\begin{equation}\label{GD2}
\frac{\partial \beta{\Pi}_\mathrm{tot}}{\partial \phi_\mathrm{tot}}=
\sum_i
\frac{\phi_i}{v_i} \frac{\partial \widetilde{\mu}_i}{\partial \phi_\mathrm{tot}}
,
\end{equation}
where $\phi_i$ and $v_i$ are the volume fraction and particle volume of component $i$, respectively, $\widetilde{\mu}_i$ is the normalized chemical potential of component $i$ and $\phi_\mathrm{tot}$ is the total volume fraction. Hence, the osmotic pressure of a multi-component mixture is given by:
\begin{equation}\label{EOSmulticomp}
\beta{\Pi}_\mathrm{tot}=\int_{0}^{\phi_\mathrm{tot}} \sum_i \frac{\phi_i}{v_i} \frac{\partial \widetilde{\mu}_i}{\partial \phi_\mathrm{tot}'} \mathrm{d} \phi_\mathrm{tot}'
,
\end{equation}
\begin{figure}[ht]
  \centering
  \includegraphics[width=0.5\linewidth]{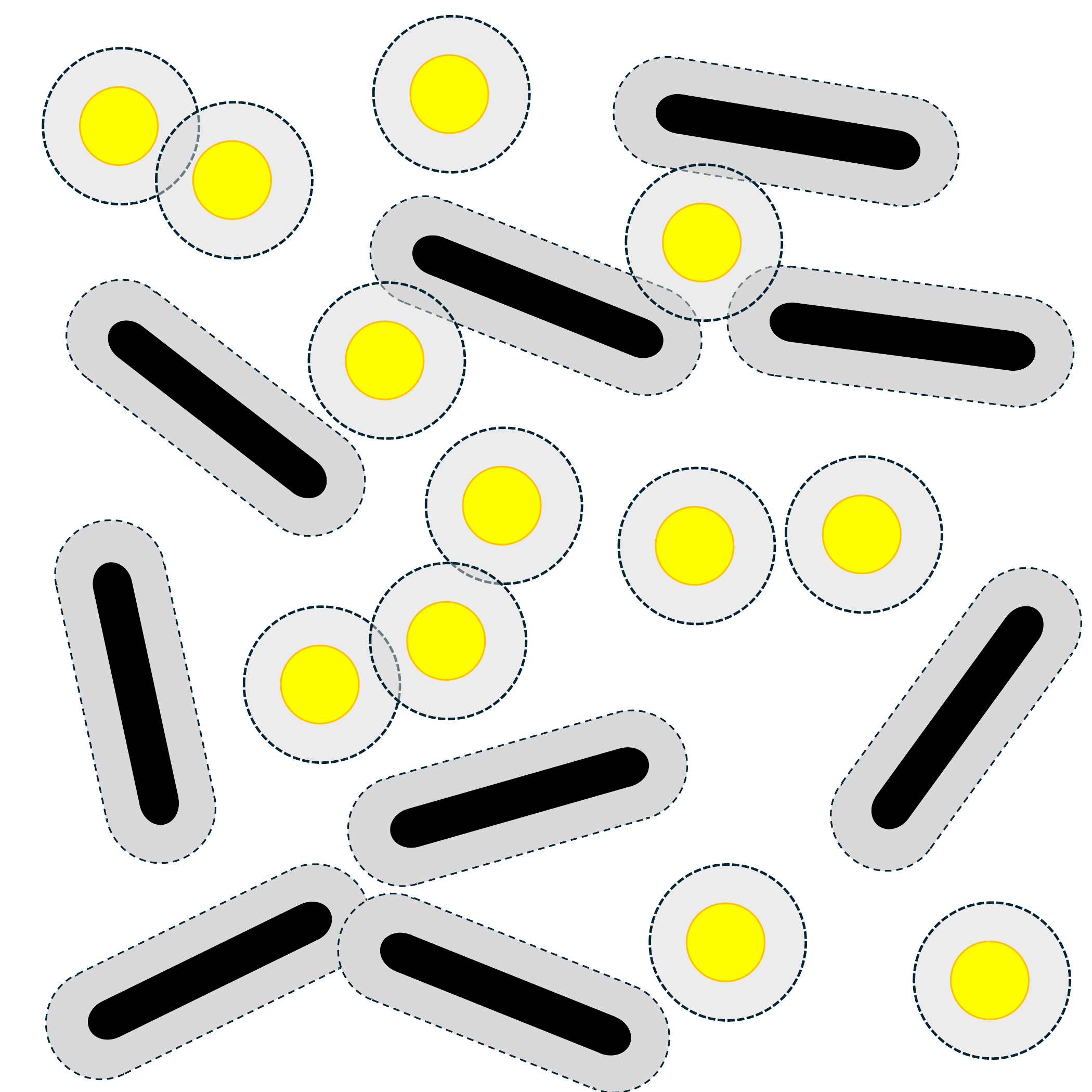}
  \caption{
Schematic representation of a binary mixture of spheres and spherocylinders in a background solvent. Dashed lines represent the excluded volume for the centres of the spheres, indicating regions inaccessible due to steric interactions.
}
  \label{fig:parttypesshapes_binary}
\end{figure}

Consider a binary mixture, such as the one sketched in Fig.~\ref{fig:parttypesshapes_binary}. The osmotic pressure of a binary mixture composed of components 1 and 2, for which $\phi_\mathrm{tot}=\phi_1+\phi_2$, follows from Eq.~(\ref{EOSmulticomp}) as:
{\begin{eqnarray}
\label{EOSbinary}
\beta{\Pi}_{\mathrm{tot}}
=
\beta{\Pi}_{1}^{1+2}
+
\int_{0}^{\phi_{2}}
\frac{\phi_{2}'}{v_2}\,
\frac{\partial \tilde{\mu}_{2}(\phi_{1},\phi_{2}')}{\partial \phi_{2}'}
\,{\rm d}\phi_{2}'\;,
\end{eqnarray}}
{where $\tilde{\mu}_{2}(\phi_{1},\phi_{2}')$ is the chemical potential of particles 2 in the binary system, and }
{\begin{eqnarray}
\label{Pi1_equation}
\beta{\Pi}_{1}^{1+2}=
\int_{0}^{\phi_{1}}
\frac{\phi_{1}'}{v_1}\,
\frac{\partial \tilde{\mu}_{1}(\phi_{1}',0)}{\partial \phi_{1}'}
\,{\rm d}\phi_{1}'
+ \frac{\phi_{1}}{v_1}
\int_{0}^{\phi_{2}}
\frac{\partial \tilde{\mu}_{1}(\phi_{1},\phi_{2}')}{\partial \phi_{2}'}
\,{\rm d}\phi_{2}'\;,
\end{eqnarray}}
{where $\tilde{\mu}_{1}(\phi_{1}',0)$ is the chemical potential of particles 1 in the absence of particles 2, and $\tilde{\mu}_{1}(\phi_{1},\phi_{2}')$ is their chemical potential in the binary system. For particles 2,} the chemical potential reads (see Eq.~(\ref{muexW})): 
\begin{equation}\label{mu2}
\widetilde{\mu}_\mathrm{2}=\widetilde{\mu}_\mathrm{2}^0 +\ln \left(\frac{\phi_2}{\alpha_\mathrm{2}^{1+2}}\right) ,
\end{equation}
where $\alpha_\mathrm{2}^{1+2}$ is the free volume fraction available for particle 2 in a mixture of 1 and 2. Combining Eqs.~(\ref{EOSbinary}) and (\ref{mu2}) gives:
{\begin{equation}\label{EOSbinaryfinal}
\beta{\Pi}_\mathrm{tot}=\beta{\Pi}_\mathrm{1}^{1+2} + \int_{0}^{\phi_\mathrm{2}}  \left(1- 
\frac{\partial \ln \alpha_{2}^{1+2}({\phi_{1},}\phi_\mathrm{2}')}{\partial \ln \phi_\mathrm{2}'} \right) \frac{\mathrm{d} \phi_\mathrm{2}' }{v_2}
.
\end{equation}}
This illustrates the importance of the free volume fraction $\alpha$: accurate expressions for $\alpha$ enable one to map out the complete (equilibrium) thermodynamic properties because they provide access to both the chemical potential as well as the osmotic pressure of particle mixtures.

\subsection{Ternary particle dispersion}\label{s:OP_tern}

\begin{figure}[ht]
\centering
\includegraphics[width=0.8\linewidth]{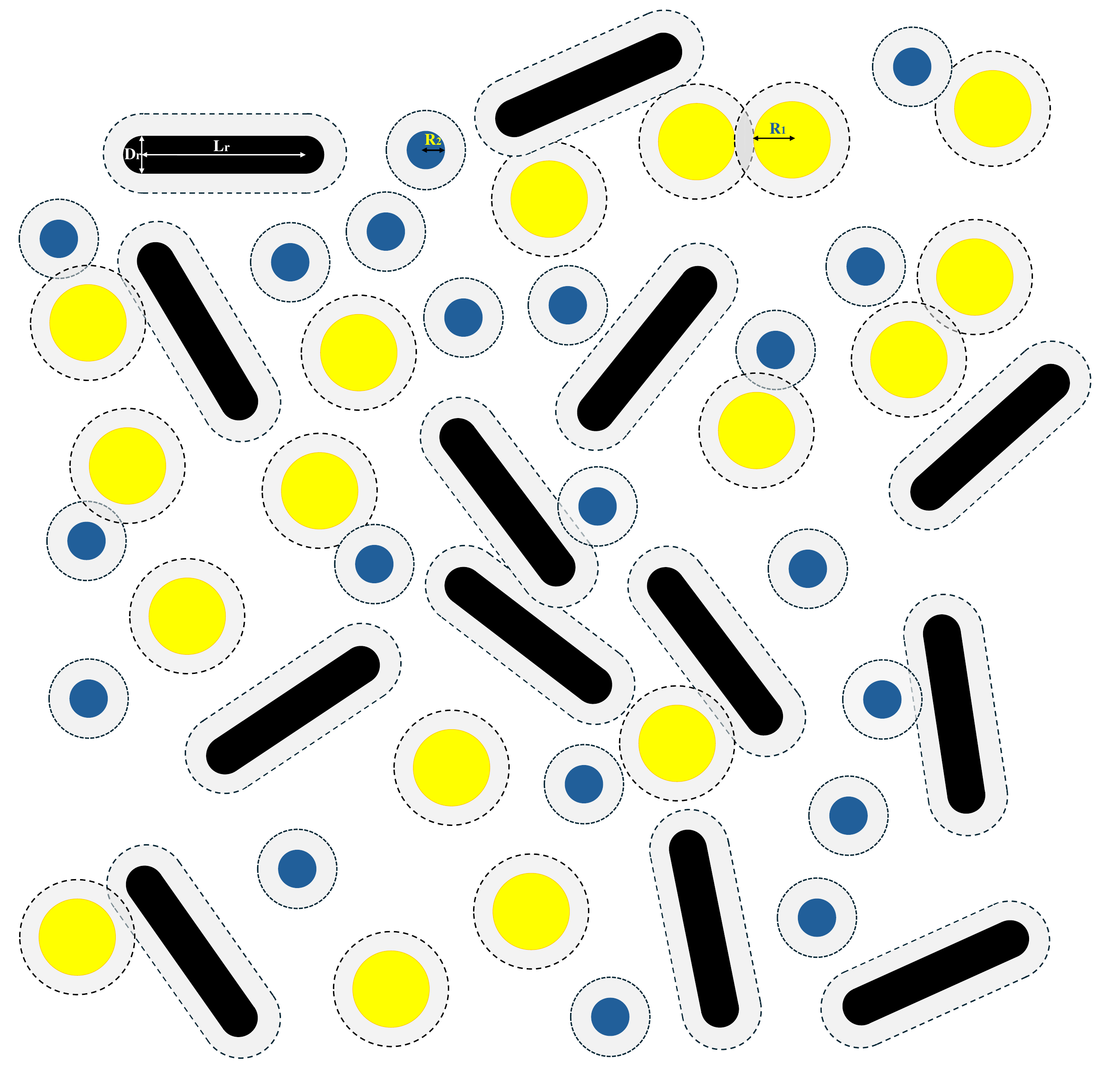} 
\caption{Ternary mixture consisting of spherocylinders (black), large spheres (yellow) and small spheres (blue). Dashed lines represent the excluded volume around each particle inaccessible to the centre of the small spheres (blue) due to steric interactions.
}
\label{fig:ternary_mixture}
\end{figure}

\noindent
We now derive an expression for the (osmotic) pressure $\Pi_\mathrm{tot}$ of a ternary mixture, starting from the general Gibbs-Duhem relation given in Eq.~(\ref{GD2}). Figure~\ref{fig:ternary_mixture} depicts an example of a ternary mixture considered here for which we derive an expression for the total osmotic pressure below. In this case, $\phi_\mathrm{tot} = \phi_1 + \phi_2 + \phi_3$ is the total volume fraction of the system. Using Widom's theorem, the excess chemical potential is expressed as $\widetilde{\mu}_i^{\mathrm{ex}} = -\ln \alpha_i{^{1+2+3}}$, where $\alpha_i{^{1+2+3}}$ represents the free volume fraction available for component $i$. Substituting this relation into Eq.~\ref{GD2}, we replace the changes in chemical potentials with the corresponding free volume fractions $\alpha_i{^{1+2+3}}$. By differentiating with respect to $\phi_3$, the normalized expression becomes:
{
   \begin{eqnarray}
\frac{\mathrm{d} \beta{\Pi}_\mathrm{tot}}{\mathrm{d} \phi_3} 
    &&= \sum_{i=1}^3  \frac{\phi_i}{v_i} \, \frac{\partial}{\partial \phi_3} \Bigl[ \ln \phi_i - \ln \alpha_i^{1+2+3} \Bigr] \nonumber \\
    &&= \frac{\phi_3}{v_3} \frac{1}{\phi_3} - \sum_{i=1}^3 \frac{\phi_i}{v_i} \left( \frac{\partial }{\partial \phi_3} \Bigl [ \ln \alpha_i^{1+2+3} \Bigr ] \right ) \nonumber \\
    &&= \frac{1}{v_3} - \sum_{i=1}^3 \frac{\phi_i}{v_i \, \alpha_i} \frac{\partial \alpha_i^{1+2+3}}{\partial \phi_3} \, .
\label{eq: d osmot / d phi_3}
\end{eqnarray}}

Now we can integrate again and find:
{\begin{eqnarray} \label{eq: osmot(_1,_2,_3)}
\beta{\Pi}_\mathrm{tot} = \beta{\Pi}^{1+2} &+&
\int_{0}^{\phi_3} \left( \frac{1}{v_3} - \sum_{i=1}^3 \frac{\phi_i'}{v_i \, \alpha_i^{1+2+3}(\phi_1,\phi_2,\phi_3')} \right. \nonumber \\
&& \phantom{\int_{0}^{\phi_3} \left( \frac{1}{v_3} - \right)}\left. \times \frac{\partial \alpha_i^{1+2+3}(\phi_1,\phi_2,\phi_3')}{\partial \phi_3'} \right) 
\mathrm{d}{\phi_3'}.
\end{eqnarray}}
Analogous to Eq.~(\ref{eq: d osmot / d phi_3}), we calculate the derivative of ${\Pi}^{{1+2}}(\phi_1,\phi_2)$ with respect {to} $\phi_2$ and the derivative of ${\Pi}^{{1}}(\phi_1)$ with respect to $\phi_1$. Then we again integrate from $\phi_2{'}=0$ {to} $\phi_2$ and $\phi_1{'}=0$ {to} $\phi_1$, respectively, and obtain the following expression:
{
{\begin{eqnarray}
\beta{\Pi}^{1+2} = \beta{\Pi}^{1} &+& \int_{0}^{\phi_2} \left( \frac{1}{v_2} - \sum_{i=1}^2 \frac{\phi_i'}{v_i\,\alpha_i^{1+2}(\phi_1,\phi_2')} \;  \right. \nonumber\\&&\phantom{\int_{0}^{\phi_3} \left( \frac{1}{v_2} - \right)}\left. \times\frac{\partial \alpha_i^{1+2}(\phi_1,\phi_2')}{\partial \phi_2'} \right) \mathrm{d}\phi_2' \,,
\label{eq: osmot(_1,_2,0)}
\end{eqnarray}}
}
with 
{{\begin{equation}
\beta{\Pi}^{1}
= \int_{0}^{\phi_1} \left( 1 -  \frac{\partial \ln \alpha_1^{1}(\phi_1')}{\partial \ln \phi_1'} \right) \frac{\mathrm{d}\phi_1'}{v_1} \,.
\label{eq: osmot(_1,0,0)}
\end{equation}}
}
\noindent where the (osmotic) pressure becomes $\Pi(0,0,0)=0$ for a pure (background) solvent in the absence of particles. Note here, that $\phi_3=0$ in Eq.~(\ref{eq: osmot(_1,_2,0)}) and $\phi_2=\phi_3=0$ in Eq.~(\ref{eq: osmot(_1,0,0)}) for all $\alpha_i$.

If the integrals are simplified by partial integration, the resulting expressions still include integral terms that cannot be calculated analytically. These remaining terms can, however, be numerically evaluated using the values of $\alpha_i$ obtained from Monte Carlo (MC) simulations for given values of $\phi_i$. By substituting Eq.~(\ref{eq: osmot(_1,0,0)}) into Eq.~(\ref{eq: osmot(_1,_2,0)}) and then substituting Eq.~(\ref{eq: osmot(_1,_2,0)}) into Eq.~(\ref{eq: osmot(_1,_2,_3)}), certain terms cancel. This simplification leads to the following expression for the total (osmotic) pressure of the ternary system:
{\begin{eqnarray}
\label{eq:Osmo_derived_JMB}
    \beta{\Pi}_\mathrm{tot} =
    &&\sum_{i=1}^3 \frac{\phi_i}{v_i} \Bigl [1-\ln \alpha_i^{1+2+3} \Bigr ]
    + \int_0^{\phi_1} \frac{\ln \alpha_1^{1}({\phi_1'})}{v_1}\mathrm{d} {\phi_1'} \nonumber \\
    && + \int_0^{\phi_2} \frac{\ln \alpha_2^{1+2}(\phi_1,{\phi_2'})}{v_2}\mathrm{d} {\phi_2'}
    + \int_0^{\phi_3} \frac{\ln \alpha_3^{1+2+3}(\phi_1,\phi_2,{\phi_3'})}{v_3}\mathrm{d} {\phi_3'} \, .
\end{eqnarray}}
This expression will be used to compute the pressure of a ternary system from simulated free volume fraction data and compared to SPT predictions for a ternary mixture.

\section{{Pressure of a ternary mixture of rods and spheres}}\label{sec:SPT_HSHSHSC}

In this section, we consider the pressure of a ternary mixture of large hard spheres, small hard spheres and hard spherocylinders, as depicted in Fig.~\ref{fig:ternary_mixture}. We first derive novel SPT results in Sec.~\ref{sebsec1:SPT_HSHSHSC}. Subsequently, we discuss Monte Carlo (MC) simulations performed in Sec.~\ref{sebsec2:SPT_HSHSHSC}, and we discuss the results and comparison of SPT and MC simulations in Sec.~\ref{sebsec3:SPT_HSHSHSC}.

\subsection{Scaled particle theory for a large +  small sphere + spherocylinder mixture}\label{sebsec1:SPT_HSHSHSC}
\begin{figure}[ht]
  \centering
  \begin{minipage}{0.2\textwidth}
    \centering
    \includegraphics[width=0.9\linewidth]{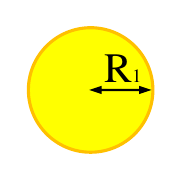} %
  \end{minipage}
  \begin{minipage}{0.3\textwidth}
    \centering
    \includegraphics[width=0.9\linewidth]{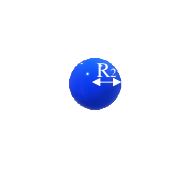} %
  \end{minipage}
  \begin{minipage}{0.4\textwidth}
    \centering
    \includegraphics[width=1.1\linewidth]{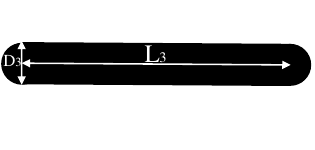} 
  \end{minipage}
  \caption{Characteristic length scales in the ternary mixture of a large hard sphere (radius $R_1$), a small hard sphere (radius $R_2$) and a spherocylinder with length $L_{3}$ and diameter $D_{3}$.
  } 
  \label{fig:system_3}
\end{figure}

We consider a ternary mixture of hard spheres (HS1 and HS2) and hard spherocylinders (HSC) with volume fractions $\phi_{\mathrm{s}1}$, $\phi_{\mathrm{s}2}$ and $\phi_3$, respectively. To characterise the particles in the mixture, we use three geometrical parameters: volume \(v\), surface area \(S\), and curvature parameter \(r\). 

 For the two types of hard spheres, HS1 with radius \(R_1\) and HS2 with radius \(R_2\), these parameters are given by the following expressions for $i=1$ and $2$:
\begin{equation} \label{eq:V1_V2_S1_S2_r1_r2}
v_{i} = \frac{4}{3} \pi R_{i}^3, \quad S_{i} = 4\pi R_{i}^2, \quad r_{i} = R_{i}
\end{equation}
where $v_{i}$, $S_{i}$, and $r_{i}$ represent the volume, surface area, and curvature parameter, respectively.

The size of a scaled particle in the case of a hard sphere (HS) below is tuned by the scaling parameter \(\lambda_{i}\). Consequently, the volume \(v^*_{{\rm }i}\), the surface area \(S^*_{{\rm }i}\), and the curvature parameter $r^*_{{\rm }i} $ of the scaled particles are given by the following expressions:
\begin{equation} \label{eq:V_is_S_is_r_is}
v^*_{i} = \frac{4}{3} \pi \lambda_i^3 R_{i}^3, \quad S^*_{i} = 4 \pi \lambda_i^2 R_{i}^2, \quad r^*_{i} = \lambda_i R_{i}
\end{equation}
where $^*$ denotes the scaled particle, and $R_{i}$ is the radius of the hard spheres (HS1 and HS2). 

We refer to particle type 3 as the hard spherocylinder (HSC). For the hard spherocylinders with diameter \(D_3\) and length \(L_3\), the parameters are:
\begin{equation} \label{eq:V3s1}\label{v3def}
v_{3} = \frac{1}{4} \pi  L_3  D_3^2 + \frac{1}{6} \pi D_3^3 ,
\end{equation}
\begin{equation}
S_{3} = \pi L_3  D_3 + \pi  D_3^2, 
\end{equation}
and
\begin{equation}
r_{3} = \frac{1}{4}  L_3 + \frac{1}{2}  D_3
\end{equation}
  
Now, we introduce the dimension of a scaled HSC particle, where \( \lambda_3 \) and \( \nu_3 \) are the scaling factors for the diameter and length, respectively:
\begin{equation} \label{D3s}
D^*_3 = \lambda_3 D_3,
\end{equation}
\begin{equation} \label{eq:L3s}
L^*_3 = \nu_3 L_3.
\end{equation}

Using these scaling parameters, the expressions for the scaled volume, surface area, and curvature parameter for the HSC particle are now given by:

\begin{equation} \label{eq:V3s}
v^*_3 = \frac{1}{4} \pi \nu_3 L_3 (\lambda_3 D_3)^2 + \frac{1}{6} \pi (\lambda_3 D_3)^3
\end{equation}

\begin{equation}
S^*_3 = \pi \nu_3 L_3 \lambda_3 D_3 + \pi (\lambda_3 D_3)^2, 
\label{eq:S_r}
\end{equation}
and
\begin{equation}
r^*_3 = \frac{1}{4} \nu_3 L_3 + \frac{1}{2} \lambda_3 D_3.
\label{eq:r_3}
\end{equation}

By inserting a scaled particle into the system, a cavity free of fluid particles is created. In the framework of Scaled Particle Theory (SPT), the excess chemical potential of the scaled particle, denoted as \(\mu^{\mathrm{ex}}_{{\rm }i}\), is calculated, corresponding to the work required to create this cavity. The expression for the excess chemical potential of small scaled particles in a mixture of hard spheres and hard spherocylinders (HS1/HS2/HSC) is given by a unified formula that takes into account the contribution of each particle type to the available free volume and the work required for insertion:
\begin{eqnarray} 
\label{eq:mu_ex}
\beta \mu^{\mathrm{ex}}_{{\rm }i}  &=& -\ln\left[1 - \phi_{\mathrm{s}1} \left(1 + \frac{r^*_{{\rm }1} S_1}{v_1} + \frac{r_1 S^*_{{\rm }1}}{v_1} + \frac{v^*_{{\rm }1}}{v_1}\right) \right. \nonumber \\
&& \phantom{-\ln\left[1\right]}- \phi_{\mathrm{s}2} \left(1 + \frac{r^*_{{\rm }2} S_2}{v_2} + \frac{r_2 S^*_{{\rm }2}}{v_2} + \frac{v^*_{{\rm }2}}{v_2}\right) \nonumber \\
&& \phantom{-\ln\left[1\right]}\left. - \phi_3 \left(1 + \frac{r^*_{{\rm }3} S_3}{v_3} + \frac{r_3 S^*_{{\rm }3}}{v_3} + \frac{v^*_{{\rm }3}}{v_3}\right)\right] .
\end{eqnarray}
For a sufficiently large scaled particle, the excess chemical potential is associated with the energy required to form a macroscopic cavity in the fluid. This energy can be described by a thermodynamic relation. For the scaled hard sphere particles, the expressions for the excess chemical potential are:
\begin{equation} \label{eq:w_lambda_1}
\beta \mu^{\mathrm{ex}}_{1} = w(\lambda_{{\rm}1}) + \beta \Pi_\mathrm{tot} v^*_{{\rm}1} =\beta W_1,
\end{equation}

\begin{equation} \label{eq:w_lambda_2}
\beta \mu^{\mathrm{ex}}_{2} = w(\lambda_{{\rm}2}) + \beta \Pi_\mathrm{tot} v^*_{{\rm}2} =\beta W_2,
\end{equation}
where \(\Pi_\mathrm{tot}\) represents the fluid pressure, and \(v^*_{{\rm}1}\) and \(v^*_{{\rm}2}\) is the volume of the scaled hard sphere. 

In the case of a scaled hard spherocylinder (HSC) particle, the excess chemical potential can be expressed as:
\begin{equation} \label{eq:w_lambda_nu}
\beta \mu^{\mathrm{ex}}_{{\rm }3} = w(\nu _{{\rm}3}, \lambda_{{\rm}3}) + \beta \Pi_\mathrm{tot} v^*_{{\rm}3} =\beta W_3.
\end{equation}

Based on the principles of SPT \cite{holovko2015, shmotolokha2022, holovko2018, holovko2020}, the functions \( w(\lambda_{{\rm}1}) \), \( w(\lambda_{{\rm}2}) \) and \( w(\nu_{{\rm}3}, \lambda_{{\rm}3}) \) can be expressed as the Taylor expansions:
\begin{equation} \label{eq:w_lambda_si_expansion1}
w(\lambda_{{\rm}1}) = \sum_{p=0}^{2} \frac{1}{p!} \left[ \frac{\partial^p w_{p}(\lambda_{{\rm}1})}{\partial \lambda_{{\rm}1}^p} \right]_{\lambda_{{\rm}1}=0} \lambda_{{\rm}1}^p, \quad 
\end{equation}

\begin{equation} \label{eq:w_lambda_si_expansion2}
w(\lambda_{{\rm}2}) = \sum_{p=0}^{2} \frac{1}{p!} \left[ \frac{\partial^p w_{p}(\lambda_{{\rm}2})}{\partial \lambda_{{\rm}2}^p} \right]_{\lambda_{{\rm}2}=0} \lambda_{{\rm}2}^p, \quad 
\end{equation}

\begin{equation} \label{eq:w_lambda_nu_expansion}
 w(\nu _{{\rm 3}}, \lambda_{{\rm 3}}) = \sum_{p=0}^{2} \sum_{q=0}^{1} \frac{1 - \delta_{p,2}\delta_{q,1}}{p!q!} \left[ \frac{\partial^{p+q}  w_{pq}(\nu _{{\rm 3}}, \lambda_{{\rm 3}})}{\partial \lambda_{{\rm 3}}^p \partial \nu_{{\rm 3}}^q} \right]_{\lambda_{{\rm 3}}=\nu_{{\rm 3}}=0} \lambda_{{\rm 3}}^p \nu_{{\rm 3}}^q.
\end{equation}

By evaluating the relevant derivatives of Eq.~(\ref{eq:mu_ex}) and substituting them into equations to Eqs.~(\ref{eq:w_lambda_si_expansion1}) (\ref{eq:w_lambda_si_expansion2}) and (\ref{eq:w_lambda_nu_expansion}), we can obtain explicit expressions for the expansion terms $w_{0}$, $w_{1}$, $w_{2}$, and $w_{00}$, $w_{10}$, $w_{20}$, $w_{01}$, $w_{11}$. Details of the derivations can be found in \ref{appendix:1}.

After setting $\lambda_{{\rm}1} = 1$ and $\lambda_{{\rm}2} = 1$ in Eqs.~(\ref{eq:w_lambda_1}) and (\ref{eq:w_lambda_2}) for the spherical particles, and $\nu_{{\rm}3} = \lambda_{{\rm}3} = 1$ in Eq.~(\ref{eq:w_lambda_nu}) for the spherocylindrical particles, we can establish a relationship between the osmotic pressure $\Pi_\mathrm{tot}$ of the mixture and the excess chemical potentials $\mu^{\mathrm{ex}}_1$, $\mu^{\mathrm{ex}}_2$, and $\mu^{\mathrm{ex}}_3$ of the fluid.

This leads to the following expressions for the chemical potentials of the individual components in the mixture:
\begin{eqnarray} \label{eq:excess_chemical_potential_1}
\beta \mu^{\mathrm{ex}}_1 &=& -\ln\left(1 - \phi_\mathrm{tot}\right) + a_1 \frac{\phi_\mathrm{tot}}{1 - \phi_\mathrm{tot}} 
+ b_1 \left( \frac{\phi_\mathrm{tot}}{1 - \phi_\mathrm{tot}} \right)^2 + \beta \Pi_\mathrm{tot}v_1 \nonumber \\
&=& - \ln {\alpha^{1+2+3}_1} ,
\end{eqnarray}
\begin{eqnarray} \label{eq:excess_chemical_potential_2}
\beta\mu^{\mathrm{ex}}_2 &=& -\ln\left(1 - \phi_\mathrm{tot}\right) + a_2 \frac{\phi_\mathrm{tot}}{1 - \phi_\mathrm{tot}} 
+ b_2 \left( \frac{\phi_\mathrm{tot}}{1 - \phi_\mathrm{tot}} \right)^2 + \beta\Pi_\mathrm{tot}v_2 \nonumber \\
&=& - \ln {\alpha^{1+2+3}_2} ,
\end{eqnarray}
\begin{eqnarray} \label{eq:excess_chemical_potential_3}
\beta \mu^{\mathrm{ex}}_3 &=& -\ln\left(1 - \phi_\mathrm{tot}\right) + a_3 \frac{\phi_\mathrm{tot}}{1 - \phi_\mathrm{tot}} 
+ b_3 \left( \frac{\phi_\mathrm{tot}}{1 - \phi_\mathrm{tot}} \right)^2 + \beta \Pi_\mathrm{tot} v_3 \nonumber \\
&=& - \ln {\alpha^{1+2+3}_3} .
\end{eqnarray}
Here, $\phi_\mathrm{tot} = \phi_1 + \phi_{\mathrm{s}2} + \phi_3$ 
is the total volume fraction of the mixture components. The coefficients $a_i$ and $b_i$ 
{are given by:}
\begin{eqnarray} 
\label{eq:a1phi}
a_1 \phi_{\mathrm{tot}} &=& 6\phi_{\mathrm{s}1} + 3\phi_{\mathrm{s}2} \frac{R_1}{R_2} + 3\phi_{\mathrm{s}2} \left(\frac{R_1}{R_2}\right)^2 
+ 12 \phi_3 \frac{\gamma}{3\gamma - 1} \frac{R_1}{D_3} \nonumber \\
&& + 12 \phi_3 \frac{\gamma + 1}{3\gamma - 1} \left(\frac{R_1}{D_3}\right)^2,
\end{eqnarray}

\begin{equation}
\label{eq:b1phi}
b_1 \phi^{2}_{\mathrm{tot}} = 0.5 \left( 3\phi_{\mathrm{s}1} + 3\phi_{\mathrm{s}2}\frac{R_1}{R_2} + 12\phi_3\frac{\gamma }{(3\gamma-1)}\frac{ R_1}{D_3} \right)^2,
\end{equation}

\begin{eqnarray} 
\label{eq:a2phi}
a_2\phi_{\mathrm{tot}} &=& 3\phi_{\mathrm{s}1}\,\frac{R_2}{R_1} + 6\phi_{\mathrm{s}2} 
+ 12\phi_3\,\frac{\gamma}{3\gamma-1}\,\frac{R_2}{D_3} \nonumber \\
&& + 3\phi_{\mathrm{s}1}\,\left(\frac{R_2}{R_1}\right)^2 
+ 12\phi_3\,\frac{\gamma+1}{3\gamma-1}\,\left(\frac{R_2}{D_3}\right)^2,
\end{eqnarray}

\begin{equation}
\label{eq:b2phi}
b_2 \phi^2_{\mathrm{tot}} = 0.5 \left( 
3\phi_{\mathrm{s}1}\,\frac{R_2}{R_1} \;+\; 
3\phi_{\mathrm{s}2} \;+\; 
12\phi_3\,\frac{\gamma}{3\gamma-1}\,\frac{R_2}{D_3} 
\right)^2,
\end{equation}

\begin{eqnarray}
\label{eq:a3phi_simpl}
a_3 \phi_{\mathrm{tot}} &=&
\frac{3}{4}\,\phi_{\mathrm{s}1}\,(\gamma + 1)\,\frac{D_3}{R_1}
\;+\;\frac{3}{4}\,\phi_{\mathrm{s}1}\,\gamma\,\left(\frac{D_3}{R_1}\right)^2
\;+\;\frac{3}{4}\,\phi_{\mathrm{s}2}\,(\gamma + 1)\,\frac{D_3}{R_2}
\;\nonumber\\[1mm]
&&+\;\frac{3}{4}\,\phi_{\mathrm{s}2}\,\gamma\,\left(\frac{D_3}{R_2}\right)^2
+
6\,\phi_3\frac{\gamma(\gamma + 1)}{3\gamma - 1}\,,
\end{eqnarray}

\begin{eqnarray}
\label{eq:b3phi}
b_3 \,\phi_{\mathrm{tot}}^2 &=&
\left(\,\frac{3}{4}\,\phi_{\mathrm{s}1}\,(\gamma - 1)\,\frac{D_3}{R_1}
\;+\;\frac{3}{4}\,\phi_{\mathrm{s}2}\,(\gamma - 1)\,\frac{D_3}{R_2}
\;+\;3\,\phi_3\,\frac{\gamma(\gamma - 1)}{3\gamma - 1}\right)
\nonumber\\[1mm]
&&\times\left(\,\frac{3}{2}\,\phi_{\mathrm{s}1}\,\frac{D_3}{R_1}
\;+\;\frac{3}{2}\,\phi_{\mathrm{s}2}\,\frac{D_3}{R_2}
\;+\;6\,\phi_3\,\frac{\gamma}{3\gamma - 1}\right)
\nonumber\\[1mm]
&&+\;0.5\;\left(\,\frac{3}{2}\,\phi_{\mathrm{s}1}\,\frac{D_3}{R_1}
\;+\;\frac{3}{2}\,\phi_{\mathrm{s}2}\,\frac{D_3}{R_2}
\;+\;6\,\phi_3\,\frac{\gamma}{3\gamma - 1}\right)^{2},
\end{eqnarray}
where $\gamma = \frac{L_3}{D_3} + 1$.

Eqs.~(\ref{eq:excess_chemical_potential_1}), ~(\ref{eq:excess_chemical_potential_2}), and ~(\ref{eq:excess_chemical_potential_3}), each of which contains two unknown quantities: one of the chemical potentials and the pressure.

In order to determine these unknowns, we will combine equations (\ref{eq:excess_chemical_potential_1}), ~(\ref{eq:excess_chemical_potential_2}), and~(\ref{eq:excess_chemical_potential_3})  with the Gibbs–Duhem relation, which was expressed for a multi-component mixture in Eq.~~(\ref{GD2}). For a 3-component mixture, the Gibbs–Duhem relation is given by:

\begin{equation}\label{GD3}
\frac{\partial \beta \Pi_\mathrm{tot}}{\partial \phi_\mathrm{tot}} = \sum_{i=1}^{3} \frac{\phi_i}{v_i} \frac{\partial \widetilde{\mu}_i}{\partial \phi_\mathrm{tot}}.
\end{equation}
By differentiating Eqs.~(\ref{eq:excess_chemical_potential_1}), ~(\ref{eq:excess_chemical_potential_2}) and ~(\ref{eq:excess_chemical_potential_3})  with respect to $\phi_\mathrm{tot}$, we obtain the following expressions:

\begin{eqnarray}\label{eq:partial_mu1}
\frac{\partial \bigl(\beta \mu_1\bigr)}{\partial \phi_\mathrm{tot}}
&=& 
\frac{1}{1 - \phi_\mathrm{tot}}
+ a_1 \,\frac{\phi_\mathrm{tot}}{(1 - \phi_\mathrm{tot})^2}
+ 2\,b_1 \,\frac{\phi_\mathrm{tot}^2}{(1 - \phi_\mathrm{tot})^3}
+ \frac{\partial \bigl(\beta \Pi_\mathrm{tot}v_1\bigr)}{\partial \phi_\mathrm{tot}}\ ,
\end{eqnarray}

\begin{eqnarray}\label{eq:partial_mu2}
\frac{\partial \bigl(\beta \mu_2\bigr)}{\partial \phi_\mathrm{tot}}
&=& 
\frac{1}{1 - \phi_\mathrm{tot}}
+ a_2 \,\frac{\phi_\mathrm{tot}}{(1 - \phi_\mathrm{tot})^2}
+ 2\,b_2 \,\frac{\phi_\mathrm{tot}^2}{(1 - \phi_\mathrm{tot})^3}
+\frac{\partial \bigl(\beta \Pi_\mathrm{tot}v_2\bigr)}{\partial \phi_\mathrm{tot}}\ ,
\end{eqnarray}

\begin{eqnarray}\label{eq:partial_mu3}
\frac{\partial \bigl(\beta \mu_3\bigr)}{\partial \phi_\mathrm{tot}}
&=& 
\frac{1}{1 - \phi_\mathrm{tot}}
+ a_2 \,\frac{\phi_\mathrm{tot}}{(1 - \phi_\mathrm{tot})^2}
+ 2\,b_2 \,\frac{\phi_\mathrm{tot}^2}{(1 - \phi_\mathrm{tot})^3}
+\frac{\partial \bigl(\beta \Pi_\mathrm{tot}v_3 \bigr)}{\partial \phi_\mathrm{tot}}\ .
\end{eqnarray}
The combination of Eqs.~(\ref{GD3}),  (\ref{eq:partial_mu1}), (\ref{eq:partial_mu2}), and  (\ref{eq:partial_mu3}) makes it possible to write down the following expression for the compressibility of the mixture:
\begin{eqnarray}\label{eq:partial_betaP}
	\frac{\partial \bigl(\beta \Pi_\mathrm{tot}v_1 \bigr)}{\partial \phi_\mathrm{tot}} = \frac{1}{1-\phi_\mathrm{tot}} + \frac{(1+A)\,\phi_\mathrm{tot}}{(1-\phi_\mathrm{tot})^2} + \frac{(A+2B)\,\phi_\mathrm{tot}^2}{(1-\phi_\mathrm{tot})^3} + \frac{2B\,\phi_\mathrm{tot}^3}{(1-\phi_\mathrm{tot})^4}\,.
\end{eqnarray}
Integration of  Eq.~(\ref{eq:partial_betaP}) provides the pressure of the ternary mixture of HS1, HS2 and HSC:
\begin{eqnarray} 
\label{eq: Pi_theory}
\beta \Pi_\mathrm{tot} v_1 &=& \left(\phi_{\mathrm{s}1} + \phi_{\mathrm{s}2} \frac{v_1}{v_2} + \phi_3 \frac{v_1}{v_3}\right) 
\left(\frac{1}{1 - \phi_\mathrm{tot}} + \frac{A}{2} \frac{\phi_\mathrm{tot}}{(1 - \phi_\mathrm{tot})^2} \right. \nonumber \\
&& \left. + \frac{2B}{3} \frac{\phi_\mathrm{tot}^2}{(1 - \phi_\mathrm{tot})^3} \right),
\end{eqnarray}
where 
\begin{equation}
A = \frac{a_1 \phi_{\mathrm{s}1}}{\phi_{\mathrm{s}1} + \phi_{\mathrm{s}2} \frac{v_1}{v_2} + \phi_3 \frac{v_1}{v_3}} + \frac{a_2 \phi_{\mathrm{s}2}}{\phi_{\mathrm{s}1} \frac{v_2}{v_1} + \phi_{\mathrm{s}2} + \phi_3 \frac{v_2}{v_3}} + \frac{a_3 \phi_3}{\phi_{\mathrm{s}1} \frac{v_3}{v_1} + \phi_{\mathrm{s}2} \frac{v_3}{v_2} + \phi_3},
\end{equation}
\begin{equation}
B = \frac{b_1 \phi_{\mathrm{s}1}}{\phi_{\mathrm{s}1} + \phi_{\mathrm{s}2} \frac{v_1}{v_2} + \phi_3 \frac{v_1}{v_3}} + \frac{b_2 \phi_{\mathrm{s}2}}{\phi_{\mathrm{s}1} \frac{v_2}{v_1} + \phi_{\mathrm{s}2} + \phi_3 \frac{v_2}{v_3}} + \frac{b_3 \phi_3}{\phi_{\mathrm{s}1} \frac{v_3}{v_1} + \phi_{\mathrm{s}2} \frac{v_3}{v_2} + \phi_3}.
\end{equation}

The last term in Eq.~(\ref{eq:w_lambda_nu}) represents the osmotic pressure, given by Eq.~(\ref{eq: Pi_theory}). This pressure times the particle volume corresponds to the work required to create a cavity for the inserted particle. By combining Eq.~(\ref{eq:w_lambda_nu}) and (\ref{eq: Pi_theory}), we ultimately obtain an expression for the work required to insert a hard spherocylinder into a ternary mixture of hard spherocylinders and hard spheres:
\begin{equation}
\beta W_3 = -\ln(1 - \phi_\mathrm{tot}) + a_3 \frac{\phi_\mathrm{tot}}{1 - \phi_\mathrm{tot}} + b_3 \left(\frac{ \phi_\mathrm{tot}}{1 - \phi_\mathrm{tot}} \right)^2 +  \beta \Pi_\mathrm{tot} v_3  ,
\end{equation}
which, using Eq.~(\ref{eq:alfaW}), directly provides the relative volume $\alpha_3{^{1+2+3}}$ available for a HSC in the multi-component mixture:
\begin{equation}
{\alpha^{1+2+3}_3} = \exp\left( - \beta W_3 \right).
\label{eq:alpha}
\end{equation}
Similarly, the work required to insert a hard sphere of type 1 (HS1) and type 2 (HS2) can be written as follows:
\begin{equation}
\beta W_1 = -\ln(1 - \phi_\mathrm{tot}) + a_1\,\frac{\phi_\mathrm{tot}}{1 - \phi_\mathrm{tot}} 
+ b_1\,\left(\frac{\phi_\mathrm{tot}}{1 - \phi_\mathrm{tot}}\right)^2 
+ \beta \Pi_\mathrm{tot}\,v_1,
\end{equation}
which directly gives the free volume fraction available for an HS1:
\begin{equation}
{\alpha^{1+2+3}_1} = \exp\left( - \beta W_1 \right).
\label{eq:alpha1}
\end{equation}
Here, $\alpha_1{^{1+2+3}}$ represents the free volume fraction available for a HS1 particle in the multi-
component mixture. Further,
\begin{equation}
\beta W_2 = -\ln(1 - \phi_\mathrm{tot}) + a_2\,\frac{\phi_\mathrm{tot}}{1 - \phi_\mathrm{tot}} 
+ b_2\,\left(\frac{\phi_\mathrm{tot}}{1 - \phi_\mathrm{tot}}\right)^2 
+ \beta \Pi_\mathrm{tot}\,v_2,
\end{equation}
 gives the free volume fraction available for an HS2:
\begin{equation}
{\alpha^{1+2+3}_2} = \exp\left( - \beta W_2 \right),
\label{eq:alpha2}
\end{equation}
where $\alpha_2{^{1+2+3}}$ is the free volume fraction available for an HS2 particle.

\subsection{Monte Carlo simulations of mixtures of large and small hard spheres and spherocylinders}
\label{sebsec2:SPT_HSHSHSC}
\begin{figure}[ht]
  \centering
  \includegraphics[trim=400 100 400 100, clip, width=0.5\linewidth]{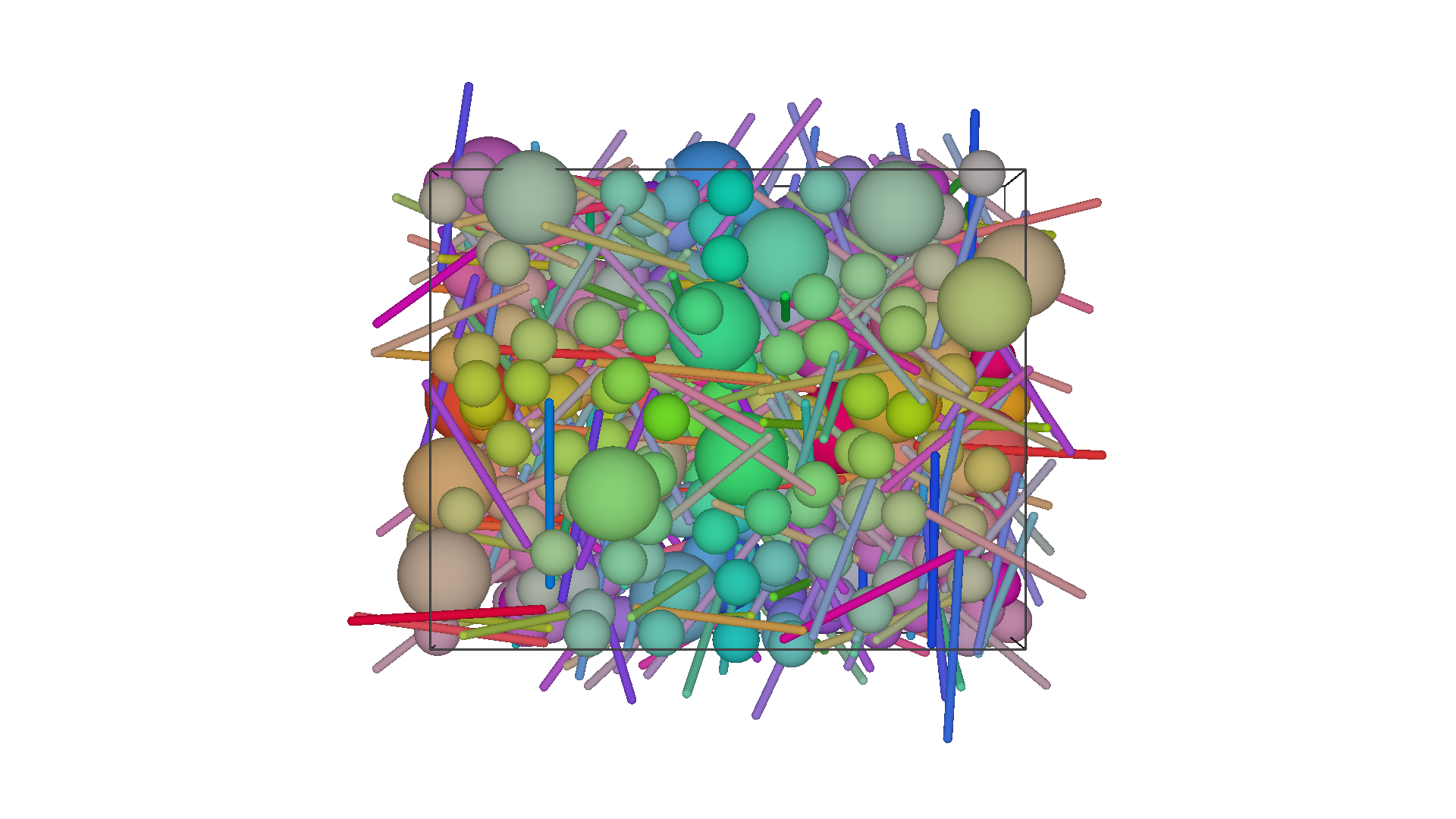}
    \caption{A snapshot of an equilibrated system of rods, large spheres and small spheres with 
  {$R_1=2.5D_3$, $R_2=5D_3$ and $L_3=20D_3$; hence $\gamma$=21, $\xi_1$=8, and $\xi_2$=4. The volume fractions of the three components are $\phi_{\rm s1}=\phi_{\rm s2}=0.1$, and $\phi_3=0.023$}.}
  \label{fig:mixture3}
\end{figure}

To verify the theoretical predictions derived in Sec.~\ref{sebsec1:SPT_HSHSHSC}, we conducted Monte Carlo simulations of {the ternary system of hard spherocylinders mixed with large and small hard spheres. An example of a simulation snapshot is presented in Fig.~\ref{fig:mixture3}.} A canonical ensemble of $N_1$ hard spheres with radius $R_1$, $N_2$ hard spheres with radius $R_2$, and $N_{3}$ hard spherocylinders (rods) with diameter $D_3$ and length $L_3$ in a cuboidal box of volume $V$ was simulated using the Metropolis Monte Carlo algorithm. As in such hard-body systems only the size-ratios matter, we measure all lengths in units of $D_3$ and use the dimensionless quantities $\gamma=L_3/D_3+1$ and $\xi_i=L_3/R_i$ to describe the systems. 

Because of the purely hard-body interaction between the particles, the Metropolis algorithm rejects any configuration in which any pair of particles overlap; all other configurations are accepted and have the same weight. In the initial setup, the rods point along a common direction, and the different species are arranged in several layers. The configurations are changed via single-particle moves, which are translations in the case of spheres and either a translation or a rotation in the case of rods. The maximal sizes of displacements and rotations are adjusted so that each type of move has an acceptance rate of about $0.35$. Most of the studied total volume fractions are low enough to result in homogeneously mixed and isotropic systems.

{To study the dependence of thermodynamic quantities such as the pressure on the volume fraction of one species while the other two have a fixed volume fraction, the number of these two particles and the box volume were fixed for a given set of parameters. The box dimensions were chosen to be at least 2.5 times the length of the rods, $L_3$, which is the largest of the length scales $(L_3, D_3, 2R_1, 2R_2)$ for the studied systems. Typical particle numbers (for the largest reported volume fractions) are $N_1:128-1797$, $N_2:16-225$, and $N_3:128-1058$. To verify results or improve insertion statistics (see below), some systems were enlarged by a factor of two or four.}

The free volume fraction of a (type of) particle is estimated from trial insertion moves. For a given configuration, a particle is put in a random position (and orientation for rods) $n_{\rm trial}$ times; if the added particle does not overlap with any of the given particles, the insertion is counted as accepted; in any case, the particle is removed after each trial. The free volume fraction is then approximately given by the acceptance rate of these trial insertions, averaged over many different configurations. 

For systems where the free volume fraction of a particle is expected to be very small, the approach of determining $\alpha$ needs an adjustment. The size of the inserted particle is initially reduced, followed by subsequent trial moves to increase it again to the original size. Sometimes, it is necessary to increase the particle size in several steps. This approach is called gradual insertion. A brief review and discussion about this adjustment can be found in Ref.~\citenum{Allen1996proc}.
The initial size and the number of intermediate sizes are automatically adjusted during the equilibration phase of the simulation{: if the number of accepted insertions is smaller than 80\% of the number of measurements, the particle size is reduced by a factor of 0.75; if the number of successful trials to increase the particle size to the original size is below 10, an intermediate size is introduced}. In between trials to change the inserted particle's size, the particle is allowed to move. 

{For each measurement configuration, at least 1000 trial insertions were attempted. This number was increased during the equilibration if the average number of successful insertions per configuration was too low (below 100). Especially for the larger sphere species, many more trial insertions ($ 10^4$-$10^6$) were required to obtain meaningful results. The number of measurements was at least $10^3$ per $10^5$ MC sweeps, where a single sweep consists of $N_1+N_2+N_3$ randomly chosen single-particle moves, so that on average each particle moves once during three sweeps. Both values were increased when the fluctuations in the average insertion acceptance became too large. Similarly, the length of the equilibration period was adjusted automatically by monitoring the slope of the average insertion acceptance 
: if the slope was larger than $\approx 5\%$ of the average value per number of measurements, the equilibration steps were doubled. After any of the mentioned adjustments, the equilibration part was continued using the new conditions.}

To determine the osmotic pressure of the ternary system as a function of the rod volume fraction, we chose the integration path {of Eq.\ (\ref{eq:Osmo_derived_JMB})} as described in {Sec.~\ref{s:OP_tern}. That means, we used a single-component} system to integrate $\ln\alpha_1{^{1}(\phi'_{\rm s1})}$ from $0$ to $\phi_{\rm s1}$, a binary system to integrate $\ln\alpha_2{^{1+2}(\phi_{\rm s1},\phi'_{\rm s2})}$ from $0$ to $\phi_{\rm s2}$, and the ternary system to integrate $\ln\alpha_3{^{1+2+3}}(\phi_{\rm s1},\phi_{\rm s2},\phi'_3)$ from $0$ to $\phi_3$. In this way, we also implicitly verified the theoretical predictions for single-type and binary particle dispersions.

\subsection{Results and Discussion}
\label{sebsec3:SPT_HSHSHSC}
In this section, the osmotic pressure of the ternary mixture determined from simulations through Eq.\ (\ref{eq:Osmo_derived_JMB}) is compared to the theoretical predictions given by Eq.~(\ref{eq: Pi_theory}). Several MC simulations were carried out for different parameter sets to analyse the osmotic pressure in relation to the volume fraction $\phi_{3}$ of the rods. {The integrals in Eq.\ (\ref{eq:Osmo_derived_JMB}) were computed} numerically using the Simpson function from the SciPy package \cite{Virtanen2019SciPy1F}. An error estimation was carried out using the Gaussian error propagation on Eq.\ (\ref{eq:Osmo_derived_JMB}), in which the standard deviations of the free volume fractions from the simulations were considered. These uncertainties of the osmotic pressure are shown as error bars in the following figures and are, if not recognisable, very small.

\subsubsection{Single particle type dispersions}
\begin{figure}[ht]
  \centering
  \includegraphics[width=0.5\linewidth]{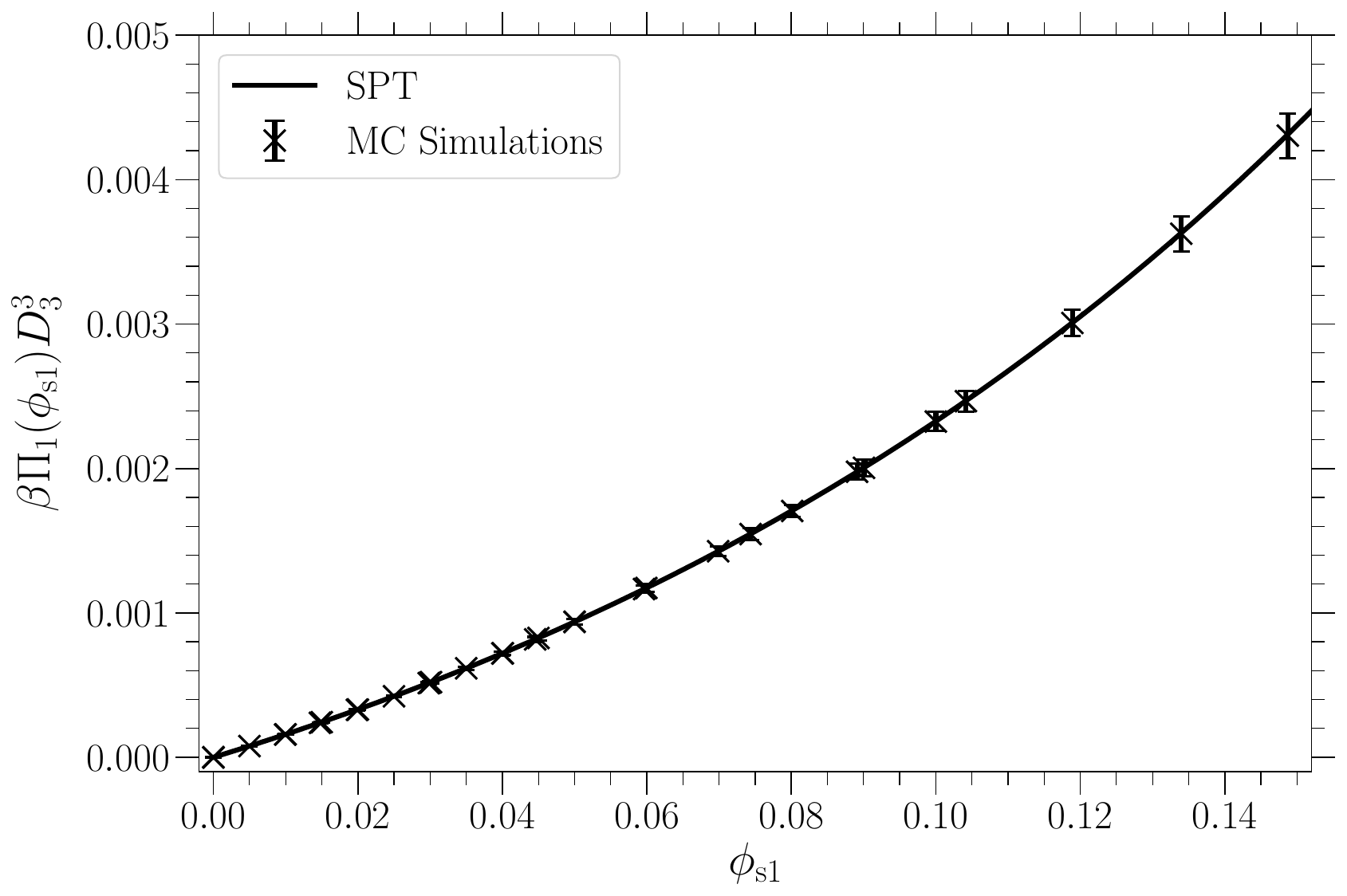}
  \caption{
  Osmotic pressure of a single-component hard sphere dispersion as a function of the particle volume fraction $\phi_{\rm s1}$. Curve follows SPT, Eq.~(\ref{eq: Pi_theory}), for $\phi_{\rm s2}=\phi_3=0$. Symbols are the results of Monte Carlo computer simulations obtained from simulated free volume fractions using Eq.\ (\ref{eq:Osmo_derived_JMB}).}
  \label{fig:pressHS}
\end{figure}

As an example of a single-component system, we show the osmotic pressure for a system of hard spheres in Fig.~\ref{fig:pressHS}, to show our MC simulations match with SPT for not too high volume fractions. The agreement between simulation results and theoretical values is almost perfect. 
This is consistent with the fact that classical Scaled Particle Theory (SPT) yields, for hard-sphere systems, the same equation of state as that obtained via the compressibility route of the Percus–Yevick (PYc) approximation \cite{PY1958,Hansen1986,Vrij2005}. SPT and PYc agree are known to be accurate up to volume fractions of approximately 40\%, beyond which notable deviations from computer simulation results begin to emerge \cite{Hansen1986,Vrij2005,Malijevski}.

\subsubsection{Ternary mixtures}
Below, we present the results for different parameter sets: (i) constant particle sizes and varying sphere volume fractions, (ii) constant sphere volume fraction and sphere diameters and varying rod lengths, (iii) constant sphere volume fractions and rod length and varying sphere diameters. 

\begin{figure}[ht]
  \centering
  \includegraphics[width=\linewidth]{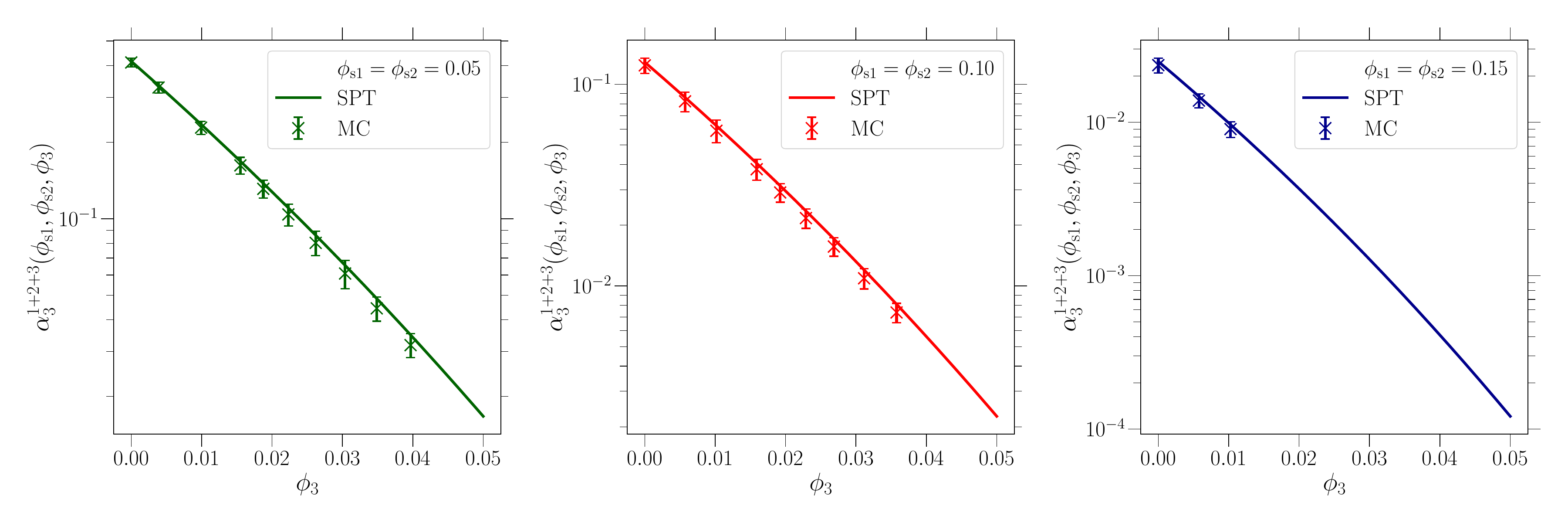}
  \caption{Free volume fraction $\alpha_{\rm 3}{^{1+2+3}}$ as a function of the rod volume fraction $\phi_{3}$. The hard sphere radii are $R_1=2.5D_3$ and $R_2=5D_3$, and the rod length is $L_3=20D_3$ {($\gamma$=21, $\xi_1$=8, and $\xi_2$=4)}. The individual graphs differ in their volume fractions of the hard sphere species 1 and 2. {The SPT predictions are given by Eq.~(\ref{eq:alpha}). {Monte Carlo simulation results are indicated by the data points. Standard deviations of these are also indicated.}}  
  }
  \label{fig:alpha_r_variing_phis}
\end{figure}

\begin{figure}[ht]
  \centering
  \includegraphics[width=\linewidth]{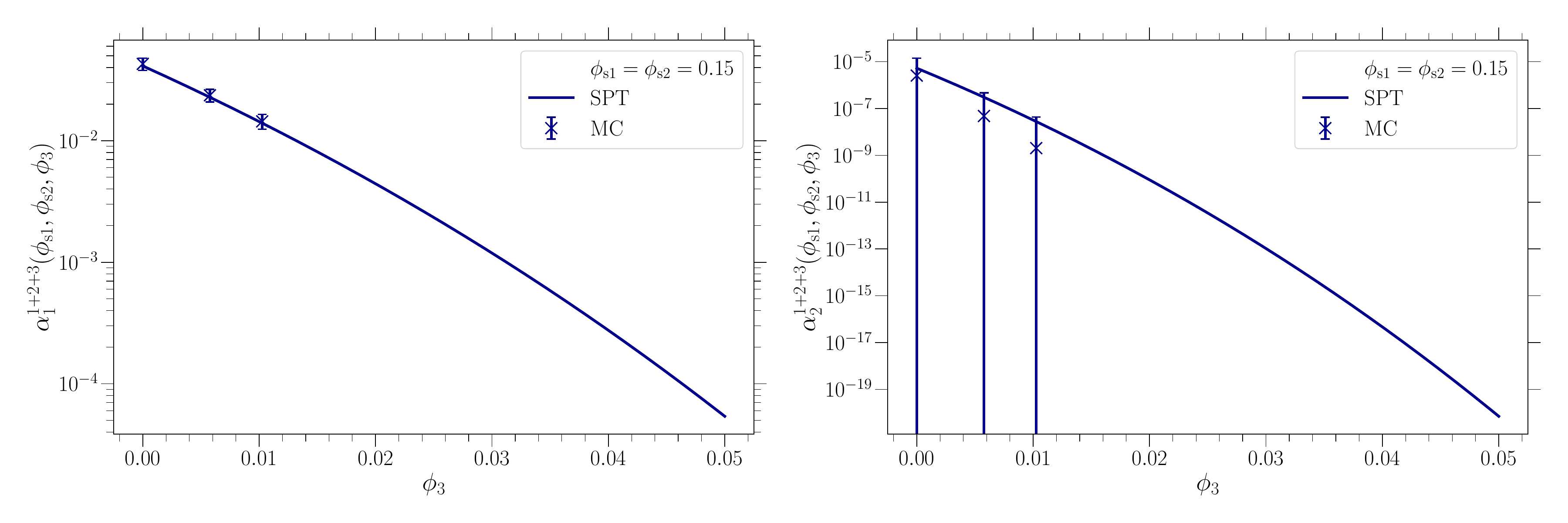}
  \caption{Free volume fractions $\alpha_1{^{1+2+3}}$ and $\alpha_2{^{1+2+3}}$ as a function of the rod volume fraction $\phi_{3}$. {Symbols refer to Monte Carlo simulation results, which include standard deviations.} As in Fig.~~\ref{fig:alpha_r_variing_phis}, the hard sphere radii are $R_1=2.5D_3$ and $R_2=5D_3$, the rod length is $L_3=20D_3$. The volume fraction of the sphere species 1 and 2 is $\phi_{\rm s1}=\phi_{\rm s2}=0.15$. {The SPT predictions are given by Eqs~(\ref{eq:alpha1}) and (\ref{eq:alpha2}). This system is an example for which the gradual insertion method was needed to estimate $\alpha_2{^{1+2+3}}$.} 
}
  \label{fig:alpha_12_phis0.15}
\end{figure}

\begin{figure}[ht]
  \centering
  \includegraphics[width=0.8\linewidth]{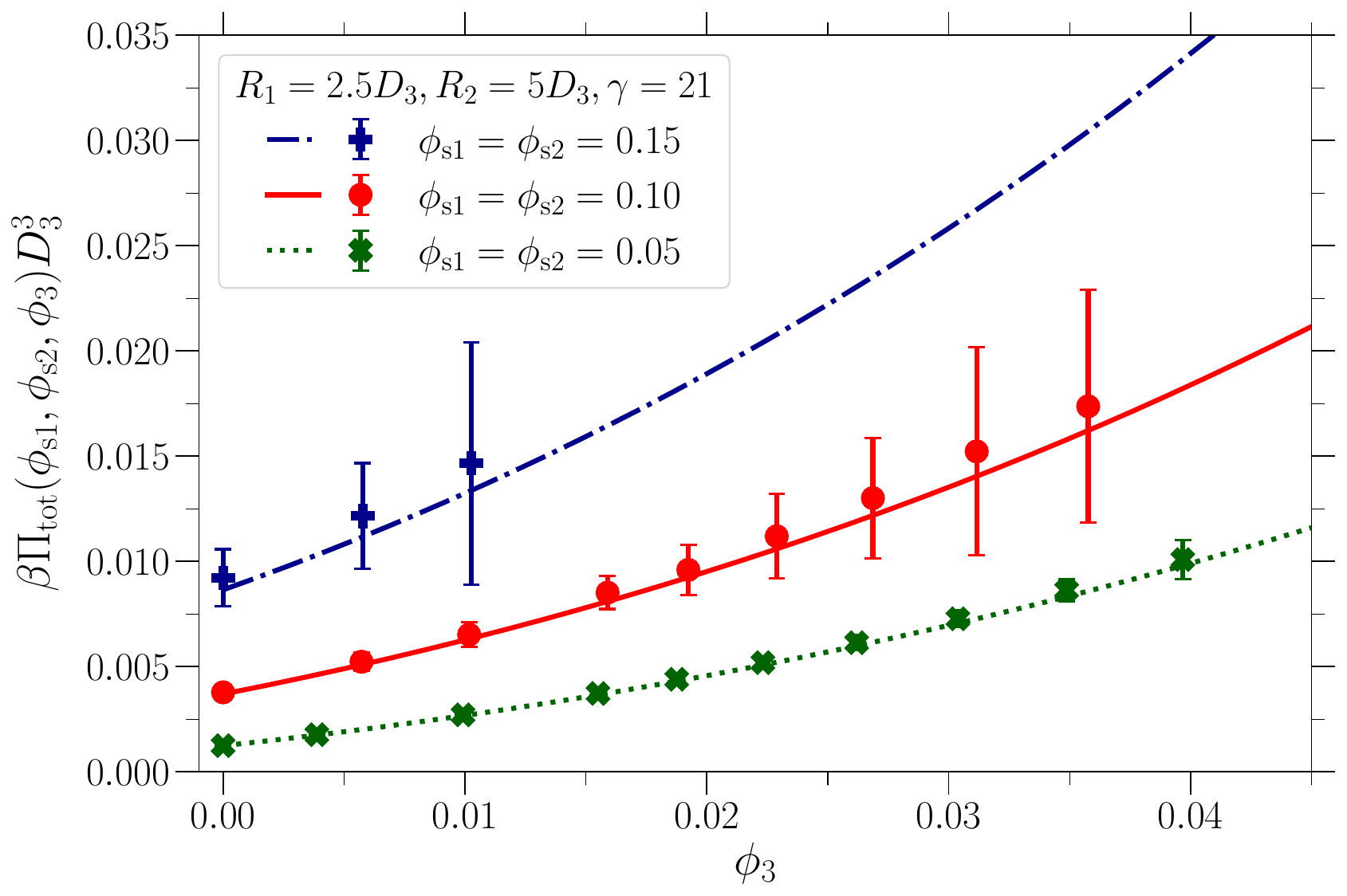}
  \caption{{Rod concentration dependence of the osmotic pressure of a mixture of small ($R_1=2.5D_3$; $\xi_1$=8) and large ($R_2=5D_3$; $\xi_2$=4) hard spheres and spherocylinders for $\gamma = 21$. The sizes of the spheres and rods with respect to each other were kept constant, while the volume fractions $\phi_{{\rm s}1}$, $\phi_{{\rm s}2}$ were varied for the three curves as indicated. Curves are SPT results following Eq.~(\ref{eq: Pi_theory}). Symbols show the results calculated from Eq.\ (\ref{eq:Osmo_derived_JMB}) using $\alpha_i$ values from computer simulations.} }
  \label{fig:mixture3_phis}
\end{figure}

In Figure \ref{fig:alpha_r_variing_phis}, the raw simulation data of {the free volume fraction available for the rods $\alpha_{\rm 3}{^{1+2+3}}(\phi_{{\rm s}1}, \phi_{{\rm s}2}, \phi_{{\rm 3}})$} used for calculating the osmotic pressure values, which are depicted in Fig.\ \ref{fig:mixture3_phis} is shown as an example. The theoretical prediction and the simulation results of the free volume fraction agree very well in the studied parameter range. For the densest shown system, only a restricted range of rod volume fractions is accessible to the simulations because the free volume fraction available for inserting hard spheres 1 ($\alpha_1{^{1+2+3}}$) and 2 ($\alpha_2{^{1+2+3}}$) is extremely small. This is shown in Fig.~\ref{fig:alpha_12_phis0.15}.

\begin{figure}[ht]
  \centering
  \includegraphics[width=0.8\linewidth]{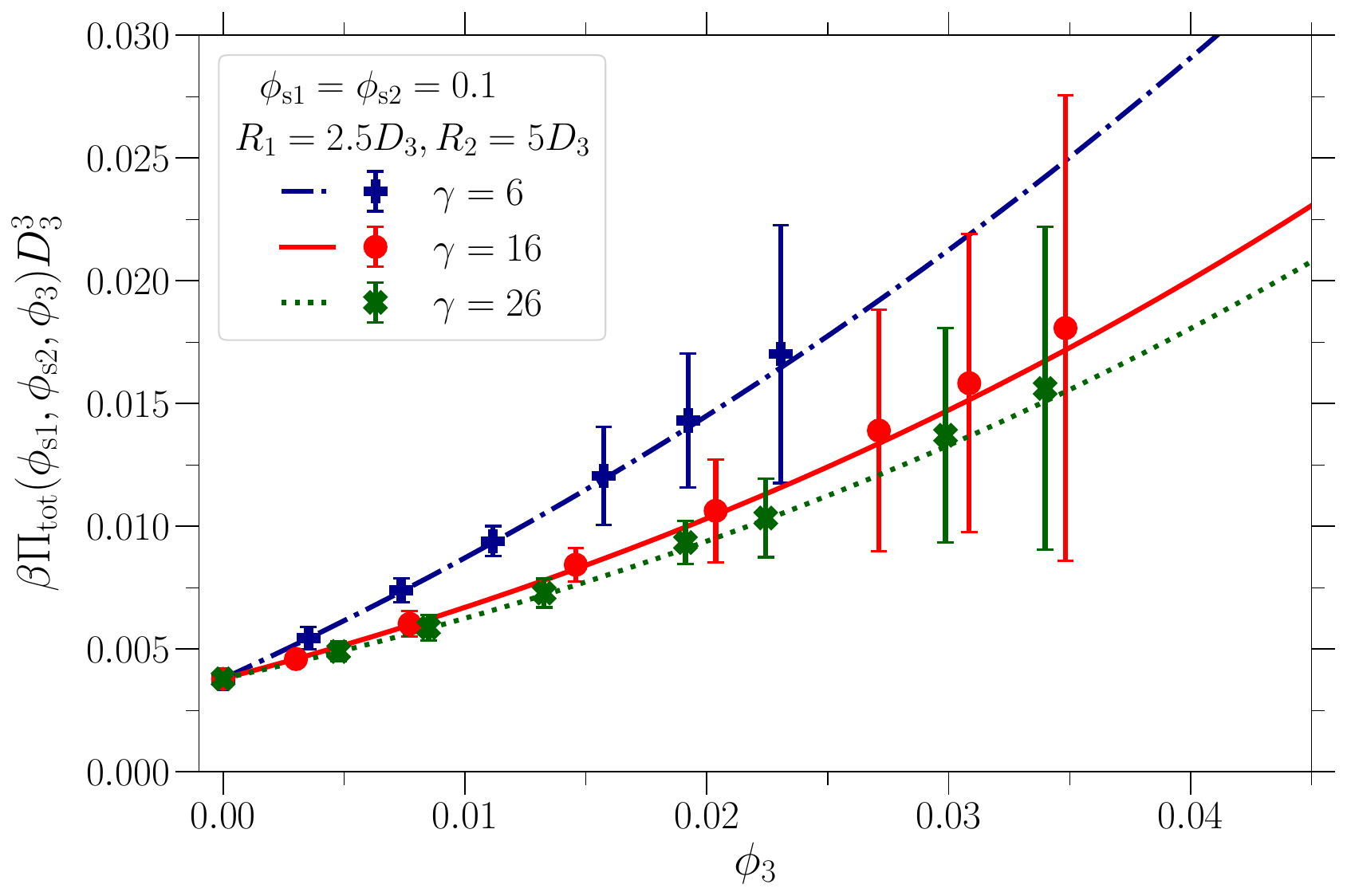}
  \caption{Osmotic pressure of a ternary system of hard spherocylinders, and small and large hard spheres as a function of the rod volume fraction for different aspect ratios (see the values of $\gamma$ as indicated), at constant size and number of spheres in the system. The curves indicate the SPT predicted values from Eq.~(\ref{eq: Pi_theory}), while the symbols indicate the calculations based on the simulation's $\alpha_i$ values using Eq.\ (\ref{eq:Osmo_derived_JMB}).
  }
  \label{fig:mixture3_AspectRartios}
\end{figure}

\begin{figure}[ht]
  \centering
  \includegraphics[width=0.8\linewidth]{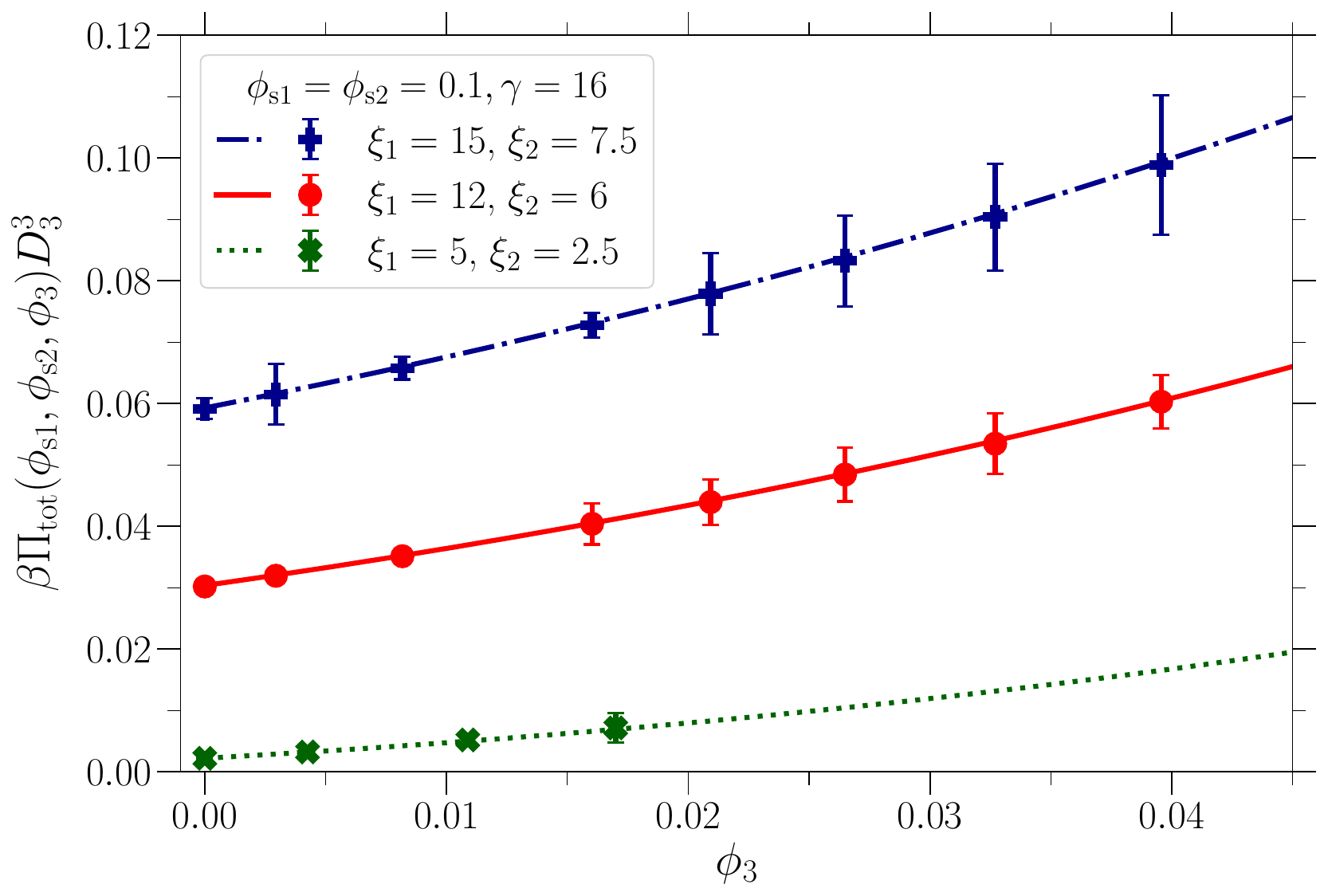}
  \caption{Osmotic pressure as a function of the rod volume fraction in a hard sphere mixture of particles 1 and 2 at fixed $\xi_1/\xi_2 = 2$ for rods with aspect ratio $L_3/D_3=15$.
 The hard sphere sizes $R_1$ and $R_2$ were varied (inset). The curves indicate the theoretical SPT predictions as in the previous plots, and the symbols again indicate the calculations based on the simulation's $\alpha_i$ values. 
  }
  \label{fig:mixture3_Xis}
\end{figure}

Figures \ref{fig:mixture3_phis}-\ref{fig:mixture3_Xis} show the results for the {rod concentration dependence of the} osmotic pressure for the sets (i)-(iii). The osmotic pressure for $\phi_{3}=0$ corresponds to the value of an asymmetric binary hard sphere system. This, in turn, includes the integration of the free volume fractions of spheres in single-particle and binary systems. As the simulation results agree very well with the theoretical predictions, we can already conclude that the theory works well for the respective one- and two-component systems and the given parameters.

As expected, the pressure increases with increasing rod and sphere volume fractions. The systems easily get dense because it is a ternary system with excluded volume interactions for the given sphere volume fractions. Therefore, the focus is on low rod concentrations. Already at low rod concentrations, the pressure increases non-linearly. In Fig.~\ref{fig:mixture3_phis} it is shown that the pressure is sensitive to the volume fraction of the spheres. 

The precise influence of anisotropic particles such as rods, in comparison to spheres, on thermodynamic properties such as pressure is nuanced. This is exemplified by Equations (7) and (12) in Ref.~[\citenum{Opdam2022}]. At a fixed volume fraction, systems composed of longer rods exhibit a lower number density, which results in a reduced osmotic pressure in the dilute regime. However, higher-order contributions to the pressure become increasingly significant with relative rod length, leading to a strong non-linear enhancement at higher concentrations.

The pressure is also larger when the rods are shorter but have the same volume fraction as long rods, cf.\ Fig.~\ref{fig:mixture3_AspectRartios}. This follows also from Eq.~(\ref{eq: Pi_theory}) from the leading $\phi_3$-term to lowest order in rod concentration, $\Pi_\mathrm{tot} \sim \phi_{3}/v_3$, but obviously also holds for higher particle concentrations. The reason is that the free volume fraction is smaller in the case of short rods. The same effect leads to an increased osmotic pressure when the spheres are {smaller} (at the same volume fraction as larger ones), cf.\ Fig.~\ref{fig:mixture3_Xis}. Of course, the pressure is sensitive to the number concentration of particles, which is larger for smaller particles at fixed volume fraction.  

Overall, theory and simulation results are in good agreement; SPT predictions are throughout within the standard deviations of the simulations that are inevitable and increase for large volume fractions. We mention that, whereas SPT is quite accurate for $L_3/D_3 \geq 5$, some deviations are expected for $1 \lesssim L_3/D_3 \lesssim 5$. For smaller rods, the Parsons-Lee approach \cite{Parsons1979,Lee1987,Lee1988} is more accurate \cite{Peters2020}.

It is noted that for high particle volume fractions (above 40 vol \%) SPT becomes less accurate. Corrections are possible \cite{Chen2016}, for instance, by replacing the SPT pressure for hard spheres with the Carnahan-Starling result \cite{Carnahan1969}. 

\clearpage
\newpage

\section{Concluding remarks}
Scaled Particle Theory (SPT) has proven to be a robust and widely applicable theoretical framework for predicting the thermodynamic properties of particle fluids. By relating these properties to the reversible work required to insert an additional particle, SPT provides a powerful yet algebraically simple approach to describing fluid systems. Since its inception by Reiss and others in the 1950s, SPT has undergone significant theoretical advancements. These include its extension to various particle geometries, such as rigid disks, spherocylinders, and superballs, as well as its adaptation for complex fluid mixtures. Cotter, Stillinger, Gibbons and others have broadened its applicability to higher-order virial corrections and structured fluid solutions.

The application of SPT to colloidal dispersions has provided insights into phase behaviour, stability, and free volume fractions within these systems. It has been applied particularly to quantify the thermodynamic properties of binary colloidal mixtures, including hard spheres, spherocylinders, and polymer additives, with predictions closely aligning with computer simulations.

While SPT is an approximate method, its predictions have been validated against computer simulations and experimental data. Refinements have improved its accuracy. The continued refinement of SPT, particularly in conjunction with Free Volume Theory (FVT), opens up new avenues for modelling phase transitions and equilibrium states in complex fluids. Future research may focus on extending SPT to include more intricate particle interactions and inhomogeneous systems, further bridging the gap between theoretical predictions and real-world applications.

We have illustrated the usefulness of the SPT concept by deriving
expressions for the pressure of a hard sphere fluid and the free
volume fraction available in such a fluid, and confirmed its accuracy by comparison with computer simulations. Subsequently, we have shown that the free volume fraction, which is a key result of
SPT, allows one to compute the chemical potential and (osmotic) pressure of a particle dispersion, and have demonstrated how this can
be extended for multi-component systems. This means that the equation of state is known if an accurate expression for the free volume
fraction is available. Hence, the free volume fraction from SPT can be used to predict the phase behaviour of (multi-component) mixtures.

Finally, we used SPT to derive the free volume fractions and osmotic pressure in a ternary mixture of hard spherocylinders and two hard sphere species. The results were tested against Monte Carlo simulations, where the free volume fractions were estimated via trial particle insertions, and the osmotic pressure was found via numerical integration from the free volume fractions. The agreement between the newly derived theoretical prediction and the simulation results is very good. In the current study, we focused on the isotropic and fluid phases of the constituent particles. Future work will also include nematic and solid phases and the determination of phase diagrams.

Based on these results, our analysis further highlights the method’s exceptional accuracy and stability in evaluating the thermodynamic properties of complex multi-component systems. Notably, this work establishes the first theoretical framework for ternary mixtures comprising hard spherocylinders and two distinct hard sphere species, thereby offering fresh insights into how geometric and size parameters influence colloidal phase behaviour. 

{
\section{Outlook}}
Future investigations are aimed at extending our study to encompass fluid-fluid \cite{Opdam2022, Opdam2023} coexistence as well as a broader range of anisotropic states, such as nematic and smectic, along with solid phases. This extended exploration is expected to deepen our understanding of phase transitions and pave the way for the development of materials with tailored adaptive properties.

Furthermore, our theoretical framework can be easily extended to elucidate cellular compartmentalisation phenomena, where the interplay between depletion forces and associative interactions drives the spontaneous formation of functionally distinct membraneless compartments \cite{Zosel2020, Speer2022}.

Moreover, future studies could focus on introducing additional parameters to account for particle interactions under high-concentration conditions more precisely. In addition, the obtained approach can serve as a basis for constructing new models to take into account van der Waals 
\cite{Franco-Melgar2009, Wu2014, holovko2015, holovko2020, shmotolokha2022} and associative interactions \cite{Hvozd2020,Hvozd2022,Shmotolokha2024CMP,Hvozd2022JCP}
for complex systems that incorporate the effects of a random porous medium \cite{holovko2018}. This may be useful in the statistical mechanical modelling of biomolecules in crowded environments \cite{Speer2022}. The planned validation of our predictions against experimental data and further numerical simulations will contribute to a more comprehensive analysis of phase transitions in these complex systems.

\clearpage
\ack
This work was financially supported by the Dutch Ministry of Education, Culture and Science (Gravity Program 024.005.020 – Interactive Polymer Materials IPM). We thank Tanja Schilling for helpful discussions regarding this project and the provision of computing resources. The support of the state of Baden-W\"urttemberg through bwHPC and the German Research Foundation (DFG) through Grant No. INST 39/963-1 FUGG (bwForCluster NEMO) is acknowledged. We thank two anonymous reviewers for many useful remarks and suggestions.

\clearpage
\appendix

\section{Taylor expansions of the immersion free energy}
\label{appendix:1}

For hard spheres, the coefficients are defined as:
\begin{eqnarray}
w(\lambda_i) &=& \sum_{p=0}^{2} \frac{1}{p!} \left[ \frac{\partial^p w(\lambda_i)}{\partial \lambda_i^p} \right]_{\lambda_i=0} \lambda_i^p. 
\end{eqnarray}
This results in the polynomial expansion up to the second order:
\begin{eqnarray}
w(\lambda_i) &=& w(0) + \left[ \frac{d w}{d \lambda_i} \right]_{\lambda_i=0} \lambda_i 
+ \frac{1}{2} \left[ \frac{d^2 w}{d \lambda_i^2} \right]_{\lambda_i=0} \lambda_i^2.
\end{eqnarray}
Thus, the coefficients are defined as:
\begin{eqnarray}
w_{p=0} &=& w_0=w(0), \nonumber \\
w_{p=1} &=& w_1= \left[\frac{\partial w}{\partial \lambda_{{\rm i}}}\right]_{\lambda_{{\rm i}}=0}, \quad \mathrm{for} \ i=1,2, \nonumber \\
w_{p=2} &=& w_2= \left[\frac{\partial^2 w}{\partial \lambda_{{\rm i}}^2}\right]_{\lambda_{{\rm i}}=0},
\end{eqnarray}
yielding the result:
\begin{eqnarray}
w(\lambda_i) &=& w_{0} + w_{1} \lambda_i + \frac{1}{2} w_{2} \lambda_i^2.
\end{eqnarray}

For hard spherocylinders, the expansion coefficients are given by:
\begin{eqnarray}
w(\nu _{{\rm 3}}, \lambda_{{\rm 3}}) = \sum_{p=0}^{2} \sum_{q=0}^{1} \frac{1 - \delta_{p,2}\delta_{q,1}}{p!q!} 
\left[ \frac{\partial^{p+q}  w(\nu _{{\rm 3}}, \lambda_{{\rm 3}})}{\partial \lambda_{{\rm 3}}^p \partial \nu_{{\rm 3}}^q} \right]_{\lambda_{{\rm 3}}=\nu_{{\rm 3}}=0}
\lambda_{{\rm 3}}^p \nu_{{\rm 3}}^q .
\end{eqnarray}
Here, the term corresponding to \( p=2 \) and \( q=1 \) is excluded by the factor \( 1-\delta_{p,2}\delta_{q,1} \). To illustrate the substitution explicitly, consider the expansion of \( w(\nu_3,\lambda_3) \). For \( p=0,\,q=0 \):
\begin{eqnarray}
w_{p=0,q=0} = w_{00}=\frac{1-\delta_{0,2}\delta_{0,1}}{0! \, 0!} w(0,0) &=& 1\cdot w(0,0) = w(0,0).
\end{eqnarray}
For \( p=1,\,q=0 \):
\begin{eqnarray}
w_{p=1,q=0} = w_{10}=\frac{1-\delta_{1,2}\delta_{0,1}}{1! \, 0!} \left[\frac{\partial w}{\partial \lambda_3}\right]_{(0,0)}\lambda_3 &=& \left[\frac{\partial w}{\partial \lambda_3}\right]_{(0,0)}\lambda_3.
\end{eqnarray}
For \( p=2,\,q=0 \):
\begin{eqnarray}
w_{p=2,q=0} = w_{20}=\frac{1-\delta_{2,2}\delta_{0,1}}{2! \, 0!} \left[\frac{\partial^2 w}{\partial \lambda_3^2}\right]_{(0,0)}\lambda_3^2 &=& \frac{1}{2}\left[\frac{\partial^2 w}{\partial \lambda_3^2}\right]_{(0,0)}\lambda_3^2 ,
\end{eqnarray}
since \( \delta_{2,2}\delta_{0,1} = 1\cdot 0 = 0 \). For \( p=0,\,q=1 \):
\begin{eqnarray}
w_{p=0,q=1} = w_{01}=\frac{1-\delta_{0,2}\delta_{1,1}}{0! \, 1!} \left[\frac{\partial w}{\partial \nu_3}\right]_{(0,0)}\nu_3 &=& \left[\frac{\partial w}{\partial \nu_3}\right]_{(0,0)}\nu_3.
\end{eqnarray}
Note that \( \delta_{0,2}\delta_{1,1} = 0\cdot 1 = 0 \). For \( p=1,\,q=1 \):
\begin{eqnarray}
w_{p=1,q=1} = w_{11} &=& \frac{1-\delta_{1,2}\delta_{1,1}}{1! \, 1!} 
\left[\frac{\partial^2 w}{\partial \lambda_3 \, \partial \nu_3}\right]_{(0,0)}\lambda_3\nu_3, \nonumber \\
&=& \left[\frac{\partial^2 w}{\partial \lambda_3 \, \partial \nu_3}\right]_{(0,0)}\lambda_3\nu_3,
\end{eqnarray}
because \( \delta_{1,2}\delta_{1,1} = 0\cdot 1 = 0 \). For \( p=2,\,q=1 \):
\begin{eqnarray}
w_{p=2,q=1} &=& w_{21}=\frac{1-\delta_{2,2}\delta_{1,1}}{2! \, 1!} \left[\frac{\partial^3 w}{\partial \lambda_3^2 \, \partial \nu_3}\right]_{(0,0)}\lambda_3^2\nu_3 \nonumber \\ &=& \frac{1-1}{2! \, 1!} \left[\frac{\partial^3 w}{\partial \lambda_3^2 \, \partial \nu_3}\right]_{(0,0)}\lambda_3^2\nu_3 = 0.
\end{eqnarray}

This provides the full result:
\begin{eqnarray}
w(\nu_3, \lambda_3) &=& w_{00} + w_{10} \lambda_3 + \frac{1}{2} w_{20} \lambda_3^2 + w_{01} \nu_3 + w_{11} \lambda_3 \nu_3.
\end{eqnarray}
These expansions serve as the basis for describing the thermodynamic properties of the components in our ternary mixture, consisting of hard spheres (HS1, HS2) and hard spherocylinders (HSC).

\section*{References}

\bibliography{JPCMref}
\bibliographystyle{iopart-num} 
\end{document}